\documentclass{article}

\usepackage{PRIMEarxiv}

\usepackage[utf8]{inputenc} 
\usepackage[T1]{fontenc}    
\usepackage{hyperref}       
\usepackage{url}            
\usepackage{booktabs}       
\usepackage{amsfonts}       
\usepackage{nicefrac}       
\usepackage{microtype}      
\usepackage{lipsum}
\usepackage{caption}
\usepackage{subcaption}
\usepackage{url}
\usepackage{fancyhdr}       
\usepackage{graphicx}       
\graphicspath{{media/}}     

\title{Physics captured by data-based methods in El Ni\~no prediction}

\author{
  G.Lancia, C.Spitoni \\
  Department of Mathematics\\
  Utrecht University, Netherlands \\
  \texttt{g.lancia@uu.nl, c.spitoni@uu.nl} \\
   \And
  I. J. Goede, H. Dijkstra \\
  Institute for Marine and Atmospheric research Utrecht, Department of Physics\\
  Utrecht University, Netherlands \\
  \texttt{i.j.goede@students.uu.nl, h.a.dijkstra@uu.nl} \\
}

\begin{document}
\maketitle

\begin{abstract}
On average once every four years, the Tropical Pacific warms considerably during events called
El Ni\~no, leading to weather disruptions over many regions on Earth. 
Recent machine-learning approaches to El Ni\~no prediction, in particular Convolutional Neural 
Networks (CNNs),  have shown a surprisingly  high skill at relatively long lead times. In 
an attempt to understand this high skill, we here use data from distorted physics simulations 
with an intermediate complexity El Ni\~no model to determine what  aspects of El Ni\~no  physics 
are represented in a specific CNN-based classification method.  We find that the CNN can adequately 
correct for distortions in the ocean adjustment processes, but that  the machine-learning method has 
far more trouble to deal with distortions in upwelling feedback strength. 
\end{abstract}


\section{ Introduction}
Interannual climate variability is strongly dominated by the El Ni\~no-Southern Oscillation 
(ENSO) in  the Tropical Pacific. During an El Ni\~no, the positive phase of ENSO, sea surface 
temperatures in the eastern Pacific increase by a few degrees 
with respect to seasonally averaged values; the oscillation phase opposite
to El Ni\~{n}o is La Ni\~{n}a, with a colder eastern Pacific. A much used  measure of the state of  ENSO 
is the NINO3.4 index, which is the area-averaged Sea Surface Temperature (SST) anomaly (i.e. 
deviation with respect to the mean seasonal cycle) over the region 170$^\circ$W-120$^\circ$W $\times$ 5$^\circ$S- 5$^\circ$N.
El Ni\~no events typically peak in December,  occur every  two to  
seven years, and their strength varies irregularly on decadal time scales. %
The spatial pattern of  ENSO variability is often determined    from  principal component 
analysis \cite[]{Preisendorfer1988}, detecting patterns of maximal variance.  At least two 
different  types of El Ni\~no events  exist \cite[]{Zhang2019}, with largest  
temperature  anomalies either  in the eastern  Pacific (Eastern  Pacific or EP El Ni\~no's) 
or near the  dateline (Central Pacific or CP El Ni\~no's).  

As ENSO has distinct  influences on the climate around  the globe  
through well-known  teleconnections, skillfull
predictions of up to a one year lead time are desired to be able 
to mitigate the effects \cite[]{balmaseda_decadal_1995}.  For ENSO 
predictions,  often the Oceanic Ni\~no Index (ONI) is used, which is 
defined as  the three-month running mean of the NINO3.4 index. 
Both statistical models (those capturing behaviour of past events) and  dynamical models 
(i.e.  those based on the underlying  physical conservation laws) are used for El Ni\~no prediction 
\cite[]{Latif1998,  Chen2008, barnston_skill_2012, Saha2014, Timmermann2018, 
Tang2018}. 
El Ni\~no events are difficult to predict as they have an irregular occurrence, and 
each time have a slightly different development \cite[]{McPhaden2015, Timmermann2018}.
Many  ENSO prediction evaluation studies  \cite[]{Barnston2012, Lheureux2017}  have shown that 
dynamical models do better than statistical models and when initialized before 
the  boreal  spring,  most models perform much worse than  when initialized in 
summer.  The latter  notion has  been indicated  by  the spring predictability barrier 
problem  \cite[]{McPhaden2003}.  

ENSO theory \cite[]{Neelin1998} provides a framework to understand the existence of 
such predictability barriers. The 
ENSO phenomenon is thought to be an internal mode of the coupled equatorial 
ocean-atmosphere system which can be self-sustained or excited by small-scale processes, often 
considered as noise
\cite[]{Fedorov2003}. Bjerknes' feedbacks are central in the amplification of SST anomalies, whereas
equatorial ocean wave processes provide a delayed negative feedback, and are responsible for  the 
time scale of ENSO.  The interactions of the internal mode and the external seasonal 
forcing can lead to chaotic behavior through nonlinear resonances \cite[]{Tziperman1994, 
Jin1994}. On the other hand, the dynamical behavior can be strongly influenced by noise, 
in particular westerly-wind bursts \cite[]{Lian2014}.  During boreal spring and summer the 
Pacific climate system  is most susceptible to perturbations leading to  predictability 
barriers \cite[]{Latif1994}. 
The growth of perturbations from a certain initial state has been investigated in detail from 
a much  used  intermediate-complexity model, the Zebiak-Cane model \cite[]{Zebiak1987}.  
Applying  the methodology of optimal modes \cite[]{Mu2007, Duan2009, Yu2012}, it  
was indeed  shown that spring is the most sensitive season for EP El Ni\~nos and likewise 
summer for CP El Ni\~nos \cite[]{Tian2015, Hou2019}.   

Deep-learning methods  (DLMs) are powerful  statistical models which have now been used in a 
wide  range of applications such as speech recognition and image reconstruction 
\cite[]{Goodfellow2016}.  These methods include   feed-forward 
Artificial  Neural  Networks  (ANNs),  Recurrent Neural Networks (RNNs),  Reservoir 
Computers (RCs), and Convolutional Neural Networks (CNNs);  over quite some time 
now,   DLMs have been applied to  El  Ni\~no  prediction  \cite{dijkstra_application_2019}.  
The current  work is motivated by the high El  Ni\~no prediction skill of two types of DLMs. 
First, in \cite[]{Ham2019} 
CNNs were trained on model data from the Climate Model Intercomparison 
Project, phase 5 (CMIP5) using transfer learning and subsequently trained on reanalysis data. The CNN-based 
scheme   shows a better forecasting skill than most  dynamical models and  this 
forecast  skill remains high up to lead times of about 17 months. It 
is  also able to successfully predict the type of El Ni\~no  (CP- or EP)  patterns which 
develop. Second, in \cite[]{Petersik2020}, Deep Ensemble Methods  \cite[]{lakshminarayanan2017simple}, 
in particular  Gaussian Density Neural Networks  (GDNNs)  and Quantile Regression Neural Networks 
(QRNNs),  were used in ENSO prediction. These methods also give  a  skillful   model for the  
long-lead time prediction  of the ONI (and its uncertainty) using a relatively small 
predictor  set. 

At the moment, there is an enormous effort to understand the performance of  DLMs 
generally referred to as explainanable AI  \cite[]{Arrieta2020}.  The research described above shows that  
DLMs are a very promising tool in  ENSO prediction that can provide useful skill of 
El Ni\~no forecasts beyond the   predictability  barriers. The intriguing question  
is now what the DLMs  capture  of the ENSO physics contained in the data.  Addressing  
this question is precisely the focus of  this paper. We will approach this issue
using the Zebiak-Cane model,  which is also  routinely used for ENSO  prediction. 
The novel aspect of this work  is that we use so-called distorted  physics experiments 
where different physical processes (such as equatorial wave dynamics and ocean-
atmosphere feedbacks) are perturbed.  Using saliency analyses, determining 
which input variables contribute most to the prediction skill,  we then aim to determine 
what part of the ENSO dynamics is represented by the DLM. 

\section{ Models and Methods}

\subsection{ ENSO model} 

The Zebiak-Cane (ZC) model \cite[]{Zebiak1987}  represents the coupled ocean-atmosphere system on an 
equatorial $\beta$-plane  in the equatorial Pacific.  In this  model, a shallow-water ocean component  
is coupled to a steady  shallow-water  Gill \cite[]{Gill1980SomeCirculation} atmosphere component.  The 
atmosphere  is driven by heat fluxes from the ocean, depending linearly on the anomaly of the sea 
surface  temperature $T$ with respect to a radiative equilibrium temperature $T_0$.  We use the 
numerically implicit fully-coupled version of this model, developed in   \cite{vanderVaart2000TheModel},  
and slightly extended in  \cite{Feng2017}. In this version, the zonal wind 
stress $\tau^x$ is written as 
\begin{eqnarray}
\label{eq:taux}
\tau^x=\tau^x_{ext}+\tau^x_c,\\
\tau^x_{ext}=-\tau_0 e^{-\frac{1}{2}\left(\frac{y}{L_a}\right)^2}.\nonumber
\end{eqnarray}
Here the external part $\tau_{ext}^x$  represents a weak ($\tau_0\sim0.01\ Pa$) easterly 
wind stress due to the Hadley  circulation, $L_a$ is the atmospheric Rossby deformation radius and 
$y$ is the meridional coordinate.  The zonal wind stress  $\tau_{c}^x$ is 
proportional to the zonal wind from the atmospheric model which, in turn, depends on sea surface 
temperature.  

As shown in \cite{vanderVaart2000TheModel}, the parameter measuring the 
strength   of all  ocean-atmosphere coupled feedbacks  is the coupling strength 
$\mu$.  When  $\mu<\mu_c$, where $\mu_c$ indicates a critical 
value, the Tropical Pacific climatology (a stationary state of the model) is stable.
However, if the coupling strength exceeds the critical value $\mu_c$, a 
supercritical Hopf bifurcation occurs and sustained oscillations occur  with a 
period of approximately four years.  A seasonal cycle is included in the model 
by varying $\mu$ over time with a specific amplitude $\Delta \mu$ and with 
an annual period. 

Apart from the coupled ocean-atmosphere 
processes, ENSO is also affected  by fast processes in   the atmosphere, 
such as westerly wind bursts.  These processes are  considered as noise in the 
ZC model. 
The representation of  atmospheric noise in the model is similar to that in \cite{Feng2017}, 
where the westerly wind bursts are represented by one Empirical Orthogonal Function 
pattern (with the associated principle component fitted to an AR(1) process) in the zonal 
wind stress. The observation-based data set in \cite{Feng2017} contains weekly patterns of 
this wind-stress noise. In the ZC model, we randomly add one of such patterns
at each time step (of a week) to the zonal wind stress. 
The effect of the noise on the model behavior depends on whether the model is in the 
super- or sub-critical regime (i.e whether $\mu$ above
or below $\mu_c$). If $\mu<\mu_c$, the noise excites the ENSO mode, causing 
irregular oscillations. In the supercritical regime, the cycle of approximately four 
years is still present, but the noise causes an irregular  amplitude of ENSO variability.

\subsection{ Distorted physics simulations} 

The advantage of the ZC  model is that the behavior of the model can be connected 
to the physical processes in a very transparent way \cite[]{Jin1997b}.  
For this model, we  define a `truth'  by a reference simulation, using an 
external seasonal cycle, prescribed noise  in the wind-stress and parameter 
settings such as in \cite{Feng2017}. Next, in subsequent distorted-physics simulations 
we change the representation of physical processes in the model by varying
parameters.  
We  will focus on the   main processes setting the time scale and amplitude  
of ENSO. 

An important memory component in the Tropical Pacific climate system is 
the ocean adjustment to changes in the atmospheric forcing. This is accomplished
by equatorial wave dynamics  and best described by a basin mode response, 
where the basin mode consists of a sum of one Kelvin and multiple Rossby waves. 
In the SST-mixed ocean dynamics mode framework behind  ENSO variability 
\cite[]{Neelin1998}, the adjustment is crucial for the timing of El Ni\~no events. 
It plays also a crucial role in the recharge/discharge oscillator view of ENSO 
\cite[]{Jin1997a}, where the equatorial heat content is varied, usually measured 
by the Warm Water Volume in observations. 
The temporal aspects of the adjustment can be controlled in the Zebiak-Cane model 
by putting a coefficient $\delta$  before the time derivatives of the ocean momentum equations 
\cite[]{Neelin1991}. In the extreme case where the time derivative is effectively zero 
($\delta = 0$), the so-called `fast-wave' limit is reached. 

Three of the most important positive Bjerknes' feedbacks  
are the  thermocline feedback, the zonal advection feedback and the upwelling (or 
Ekman) feedback \cite[]{DijkstraB2005}. The relative magnitude of these feedbacks 
determines which spatial SST perturbation patterns are amplified. In addition, the 
feedbacks  determine also the mean state and seasonal cycle of the tropical Pacific 
climate state \cite[]{Dijkstra1995}. 
Specific feedback strengths can be 
changed in the ZC model by varying the  mean thermocline depth (thermocline feedback), the 
mean zonal temperature  gradient (zonal advection feedback), or Ekman friction 
(upwelling feedback). We will concentrate on the latter feedback, affecting the 
amplitude of ENSO,  and which can be changed in the ZC model by adjusting 
the parameter $\delta_s$. 

\subsection{ CNN approach} 
Due to their versatility and peculiarity in solving binary and multi-labels classification tasks by capturing and recognizing the discerning patterns of the input data, CNNs (Convolutional Neural Networks) can represent a powerful method for making forecasting of ENSO events with lead times of up to one and a half years \cite[]{ham2019deep} or for solving a binary classification problem in hybrid models with high complexity multi-resolution input data \cite[]{yan2020temporal}.
Unlike more sophisticated and popular ANNs like CNN-LSTM and ConvLSTM, the predictions provided by the CNN can be made explainable by means of saliency maps
\cite[]{adebayo2018sanity,selvaraju2017grad, zhou2016learning, montavon2019layer, mundhenk2019efficient} that allow to outline the spatial locations of those signal patterns that mainly contribute to make the CNN give the classes of output.
Therefore, CNNs represent the perfect choice for classifying the occurrence of ENSO events in ZC simulations and investigating in detail on which features contained in data can lead to highly accurate predictions.
To leverage the basic feature of the CNN of encoding the sequentiality of the patterns contained in the input data, we feed the CNN with simulated time series obtained via Zebiak-Cane model.
This synthetic data set describes the temporal evolution in the NINO3.4  region of some physical observables of interest as the thermocline depth, the sea surface temperature, the wind speed and the wind stress noise. 
The extraction of the instances from the ZC simulations therefore consists in slicing the synthetic time series along the time-domain, i.e. the set of time series is chunked in a sequence of overlapping time windows of 48 months and stride 1.
As a result, each single instance is a tensor of rank 2 whose dimensions are the time-length (48 pixels, sampling frequency one month) and the number of time series features (the four physical observables of interest).

\begin{figure}[!] 
\begin{subfigure}{.2\linewidth}
    \includegraphics[scale=.2]{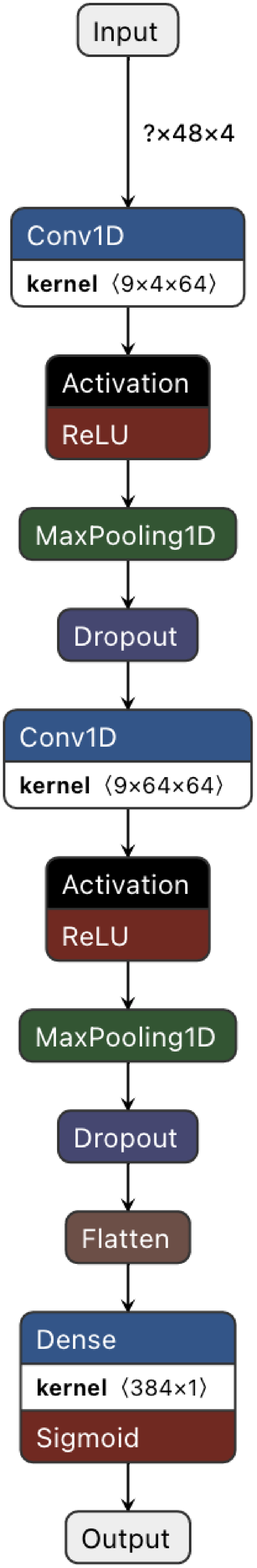} \\
    \caption{Flow diagram}
  \end{subfigure} 
  \begin{subfigure}{.8\linewidth}
    \includegraphics[width=1\linewidth]{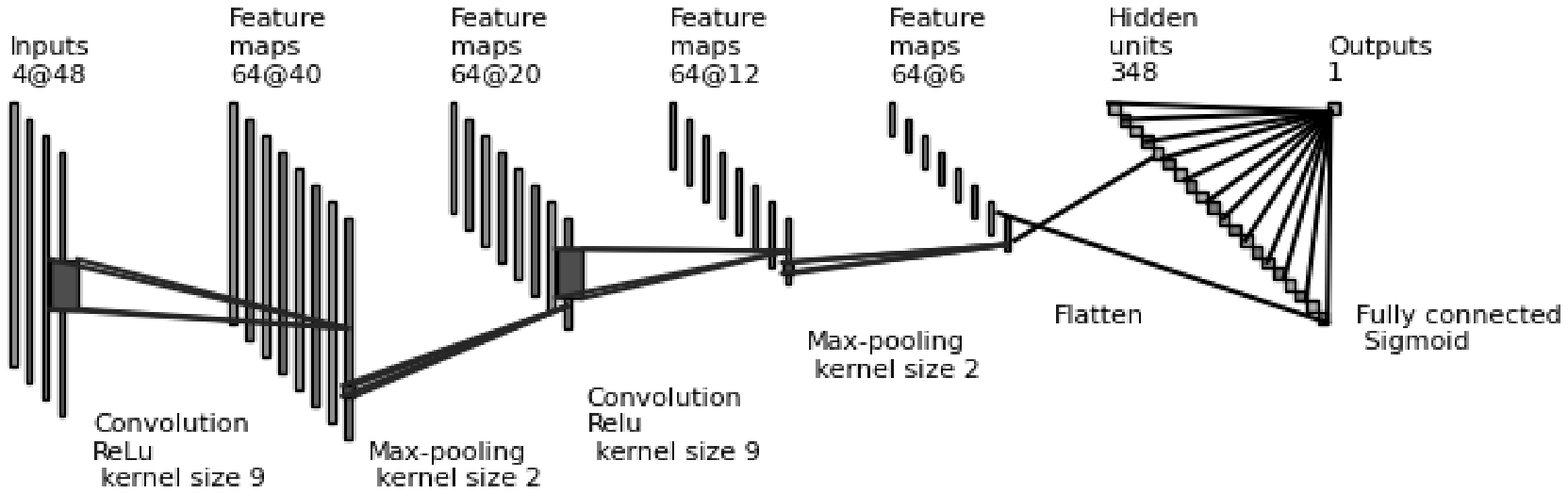} \\
    \caption{Hidden layers}
  \end{subfigure} 
    \caption{Schematic illustrations of the CNN model}
    \label{fig:CNN scheme}
\end{figure}

The labeling of the instances is performed by equipping each instance with the corresponding ONI-index value, and so we label one ENSO event whenever the ONI-index value is grater than 0.5 (El Ni\~no event) or lower than -0.5 (La Ni\~na event). 
The inputs instances are then pre-processed via standardization (each feature has now zero-mean and unit variance) and divided into training set 
and test set; we validate the CNN model by means of the 5-fold cross-validation.
Therefore we evaluate the AUC (Area Under the Curve) of the receiver operating characteristic curve on each fold; the mean value and the standard error mean will provide the degree of accuracy of the CNN model and its error respectively. 
The design of our CNN is quite standard  and it is composed by the sequence of one convolutional layer (64 kernels, size 9) with a Rectified Linear Unit (ReLu) activation function, followed by a maxpooling Layer with pooling size 2 (Fig.~ \ref{fig:CNN scheme}). 
Dropout layers \cite[]{srivastava2014dropout} with dropout rate 0.50 are also employed to reduce overfitting, but no stride is applied during the convolutions.
After repeating two times this block of hidden layers  the resulting feature map is flattened via a flatten layer; the final fully-connected layer with sigmoid activation function returns the output of the CNN.
During the training phase the ADAM \cite[]{kingma_adam:_2014} algorithm is used as optimizer for the Binary-Cross Entropy loss function; the batch size and learning rate are set equal to 128 and 0.005,  respectively.
SMOE scale method \cite[]{mundhenk2019efficient} is a novel and robust measure statistics of the activation values of CNNs arising at different spatial locations (temporal locations in the domain of our instances). 
This statistics can be used to construct robust saliency maps that appears to be much more efficient and computationally faster than popular gradient methods.
We therefore exploit the capability of SMOE scale method to detect those patterns and their spatial domains that mostly indicate the approaching or the occurring of the ENSO events.
Thus, we proceed with the analysis of the profile of the saliency maps in order to evince possible analogies and differences between the patterns learnt during the training phase and the patterns contained in the test data set.
In order to complete the analysis provided by the SMOE scale method, we even look at how the predictions can change when 
only a spectral sub-band of the input instances is propagated through the hidden layers.

With this approach we aim to investigate how oscillations occurring under a specific regime can really be a basic 
aspect of the prediction provided by the CNN. 
Therefore, we can progressively apply a digital Butterworth (\cite{butterworth1930theory}, \cite{butter_hamming1998digital}) filter of
order 3 as either a bandpass filter or low-pass filter to smooth the input instance. 
The ensemble of bandpass filters are designed to cover the whole spectral domain of any input instance and be non-overlapping at the same time and thus we impose the cutoff frequencies of each filter to be in ratio 1:2. 
This means that, starting from the Nyquist frequency $\nu_0$,  the first digital filter will have its frequency band in $[\frac{\nu_0}{2}, \nu_0]$, the second one in $[\frac{\nu_0}{4}, \frac{\nu_0}{2}]$, and so on. 
Again, when considering the low-pass 
digital filters we will choose the cut-off frequency according to a dyadic scale, i.e. the first filter will have cut-off frequency $\nu_0$, the second one $\frac{\nu_0}{2}$, the third one $\frac{\nu_0}{4}$ and so on.
\begin{table}
        \caption{Frequency bands and cut-off frequencies for the band-pass and low-pass digital filters,  respectively}
        
        \begin{tabular}{|c||c|}
        \hline
        Frequency bands (Period in months)  & Cut-off period (months)\\
        \hline
        $[2, 4)$ & 2\\ \hline
        $[4, 8)$ & 4\\ \hline
        $[8, 16)$ & 8\\ \hline
        $[16, 32)$ & 16\\ \hline
        $[32, 48)$ & 32\\ \hline
        \end{tabular}
        \label{tab:bandwiths}
    \end{table}
The full list of bandwidths (in periods) and cutoff frequencies is reported in Table~\ref{tab:bandwiths}.
Note that we will apply these digital filtering techniques by repeating the same 5-fold cross validation with metrics AUC, as we do in the model validation; the CNN architecture will not be altered during this step.
Hence, by means of this approach we aim to reveal which time scale is dominant in those patterns that characterize 
the ENSO events (e.g. a slow oscillating trends against rapid oscillating deviations), i.e. we make an effort to understand how the periodicity of the time series features is an essential characteristic of data that the CNN captures for solving the classification task and how a distortion of it can give rise to a decrease in the CNN capability of classifying the El Ni\~no and La Ni\~na events.  

\section{ Results} 

\subsection{ Distorted physics}
The model experiments broadly consist of two steps:  first the ZC model is run for standard parameter
values to produce reference case data. Then it is ran again but for a range of values around 
the standard parameter value (shown in Table~\ref{table:parameters}) to get the distorted data. 
This ultimately results in three 
different kinds of datasets: reference case, 
distorted wave speed and  distorted upwelling feedback. There are  no simulations 
where more than one parameter is distorted at the same time.  
In the second step,  the distorted datasets are used as training data for the DLMs whose 
performance is then determined by using the reference case as the test set. As a consistency 
check the DLMs are also trained on the  reference case data and then tested on reference case data. 
This should produce the highest performance  because the DLMs are tested on data they have already seen. 
\begin{table}[!]
\caption{Parameter settings of the ZC model  used to generate the data used in the 
distorted physics experiments with the parameter step size shown within brackets. Parameter 
ranges are chosen to cover roughly a 50\% increase and decrease compared to reference value, 
step size is chosen to get around 10 points within this range. The parameters are from left to right: 
coupling strength $\mu$, wave speed parameter $\delta$ and upwelling feedback parameter 
$\delta_s$. The value of $\mu = 2.7$ is subcritical in the ZC model. }
\begin{tabular}{|l||c|c|c|c|c| }
\hline 
effect & $\mu$ & $\delta$ & $\delta_s$ & $\delta_{sst}$ \\ \hline \hline
distorted wave speed & 2.7 & 0.5-1.5 (0.1) & 0.3 & 1.0  \\ \hline
distorted wave speed & 2.7 & 0.5-1.5 (0.1) & 0.3 & 1.0  \\ \hline
distorted upwelling & 2.7 & 1.0 & 0.1-0.6 (0.05) & 1.0  \\ \hline
distorted upwelling & 2.7 & 1.0 & 0.1-0.6 (0.05) & 1.0  \\ \hline
reference  & 2.7 & 1.0 & 0.3 & 1.0  \\ \hline
\end{tabular}
\label{table:parameters}
\end{table}

\subsection{Equatorial wave dynamics} 

\begin{figure*}[!]
    \includegraphics[width=1.\linewidth]{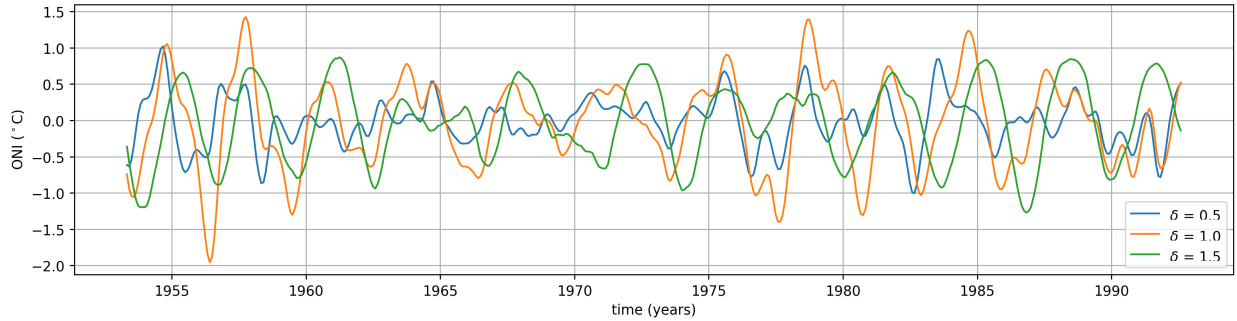}
    \caption{Several  time series  of  ONI  calculated  from  ZC model  simulations  using $\delta$ parameter  
    values  of  0.5,  10  and  1.5, respectively. }
    \label{fig:ONI_for_3_delta_values}
\end{figure*}
Time series of the ONI for the different $\delta$ values, as computed from the ZC model are
shown in Fig.~\ref{fig:ONI_for_3_delta_values}. Changing the $\delta$ value  causes  the amplitude of 
the oscillation to become much smaller for $\delta < 1$, so much even that by definition only ENSO 
neutral conditions ($-0.5 < ONI < 0.5$) are present. Increasing $\delta$ above the reference value of 
1.0 initially leads to an increase in the oscillation amplitude and it then decreases again for higher values 
of $\delta$. This is expected because the  ENSO period depends on the speed of Rossby and Kelvin 
waves crossing the Pacific basin. In the study of the classification performance of the CNN, we take 
a prediction lead time of 9 months. 
\begin{figure}[!] 
\begin{subfigure}{0.5\linewidth}
    \includegraphics[width=1\linewidth]{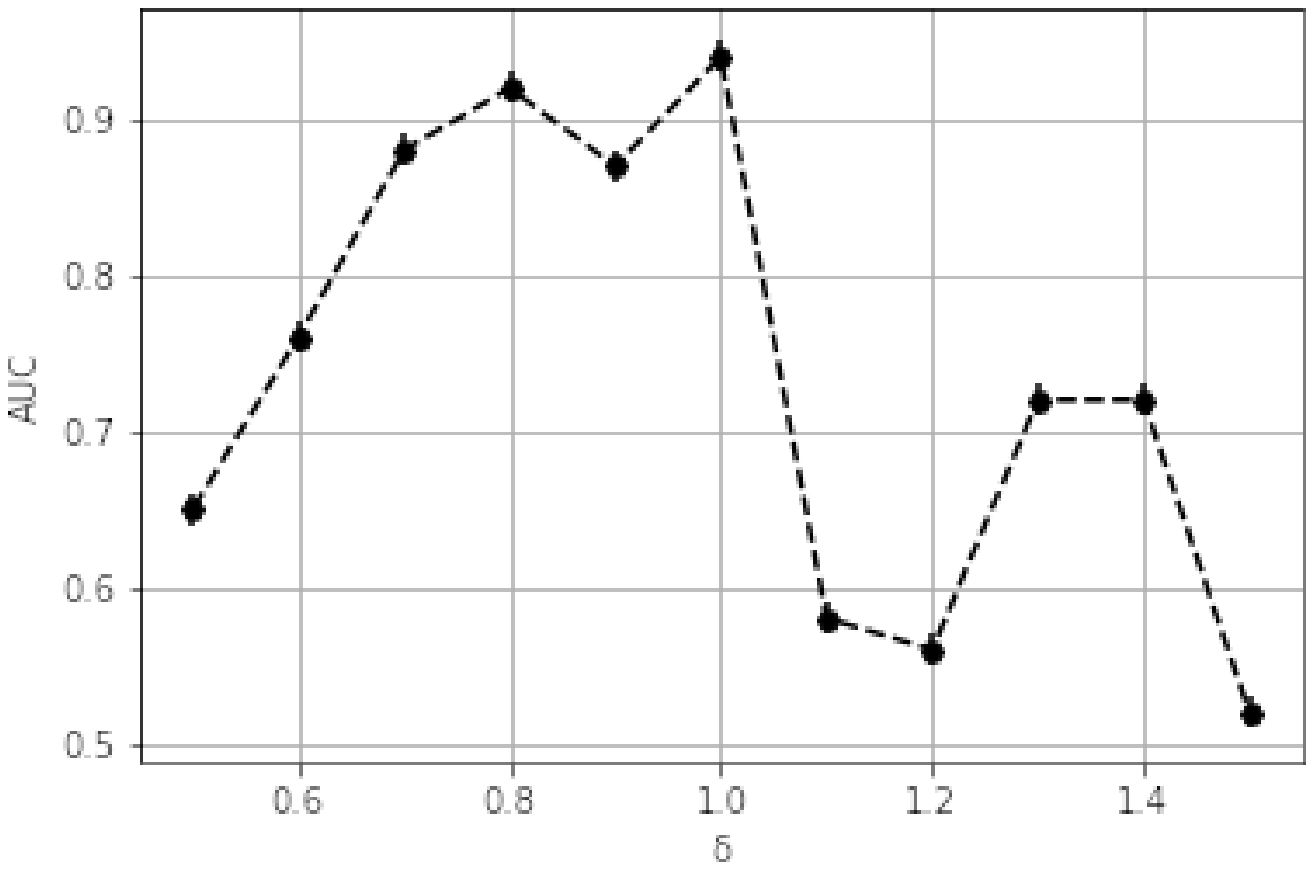} 
    \caption{}
  \end{subfigure} 
  \begin{subfigure}{0.5\linewidth}
    \includegraphics[width=1\linewidth]{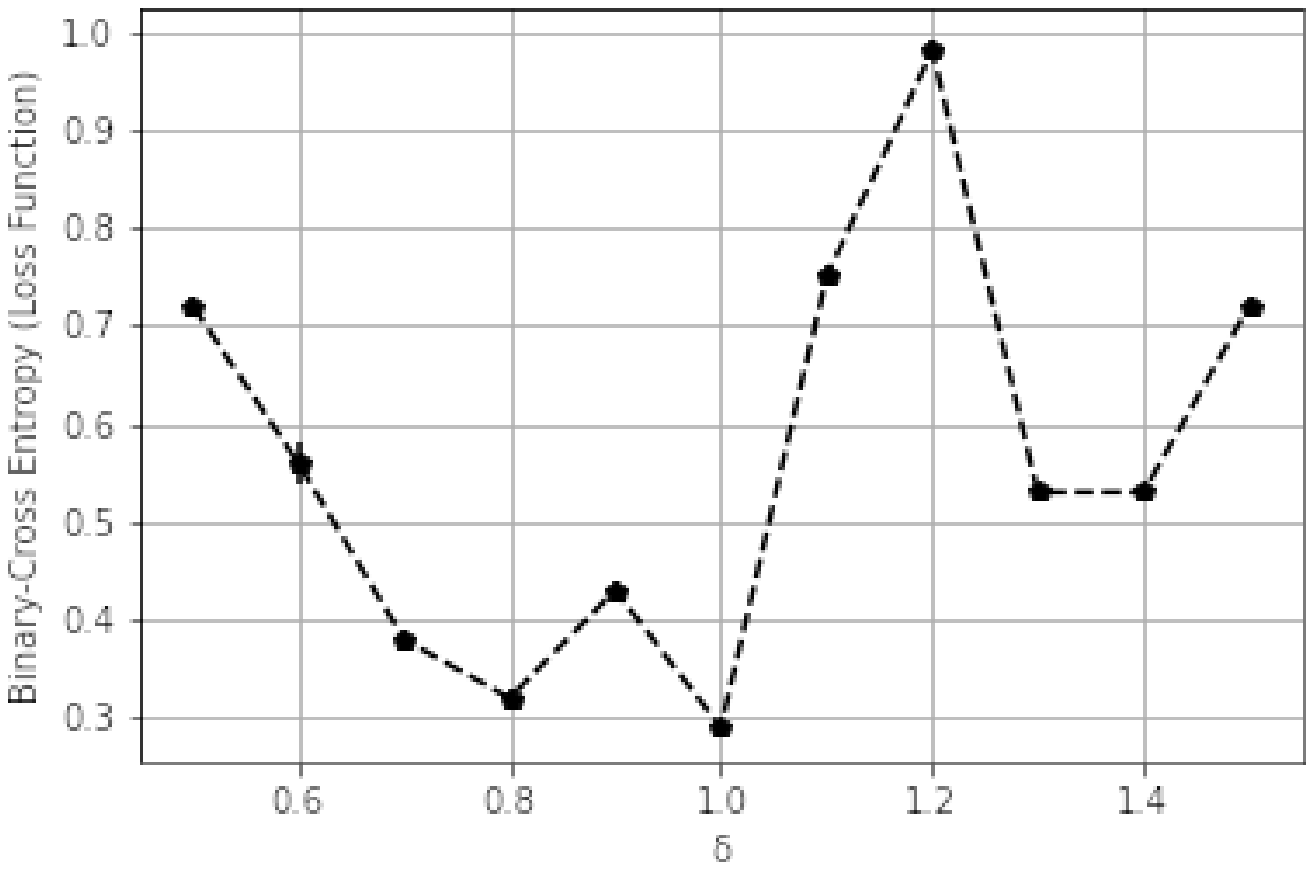}
    \caption{}
  \end{subfigure} 
  \caption{The AUC score (a) and the loss function (b) as a function of the equatorial wave speed $\delta$. Each point represents the mean AUC over 5 different folds; error bars are evaluated via standard error mean.}
   \label{fig:auc_delta} 
\end{figure}
The propagation of the $\delta$-distorted data through the CNN can lead to substantial 
changes when testing the accuracy of model on the reference data.  By construction, the 
AUC score (Fig.~\ref{fig:auc_delta}) attains excellent results at $\delta = 1.0$ (AUC 0.94) 
as the CNN is trained on the reference data.  The AUC scores tend to remain relatively high  
(peak of AUC 0.91 at $\delta= 0.8$)
as the $\delta$ parameter is slightly deceased from its reference value.
Instead, as $\delta$ is reduced up to value $0.5$ we can observe a severe degradation of the accuracy with 
respect to the reference case; from $\delta= 0.7$ the evaluation of the AUC metrics decreases monotonically 
(AUC 0.66 at $\delta= 0.5$). At values $\delta > 1.0$ we observe a total reduction of the AUC values. 
Specifically, models trained for  $\delta 
= 1.1$, and $\delta = 1.2$ show low AUCs as 0.58 and 0.56 but  the lowest value (AUC value 0.51) is reached 
at $\delta= 1.5$. 
The evaluation of the loss function (when the reference data are propagated through the CNN models) confirms the scenario expressed above.
Indeed, the global minimum value is achieved at $\delta = 1.0$ and a relative minimum is also present at $\delta = 0.8$.   
When $\delta$ is decreased or augmented towards the bound values $\delta = 0.5$ and $\delta = 1.5$ respectively, we can observe the loss function tends to reach higher values.
In particular, an increase or decrease in the AUC along the $\delta$ domain is followed by a decrease or an increase in the loss function.

\begin{figure}[!]
    \begin{subfigure}{0.4\linewidth}
    \centering
    \includegraphics[scale= .35]{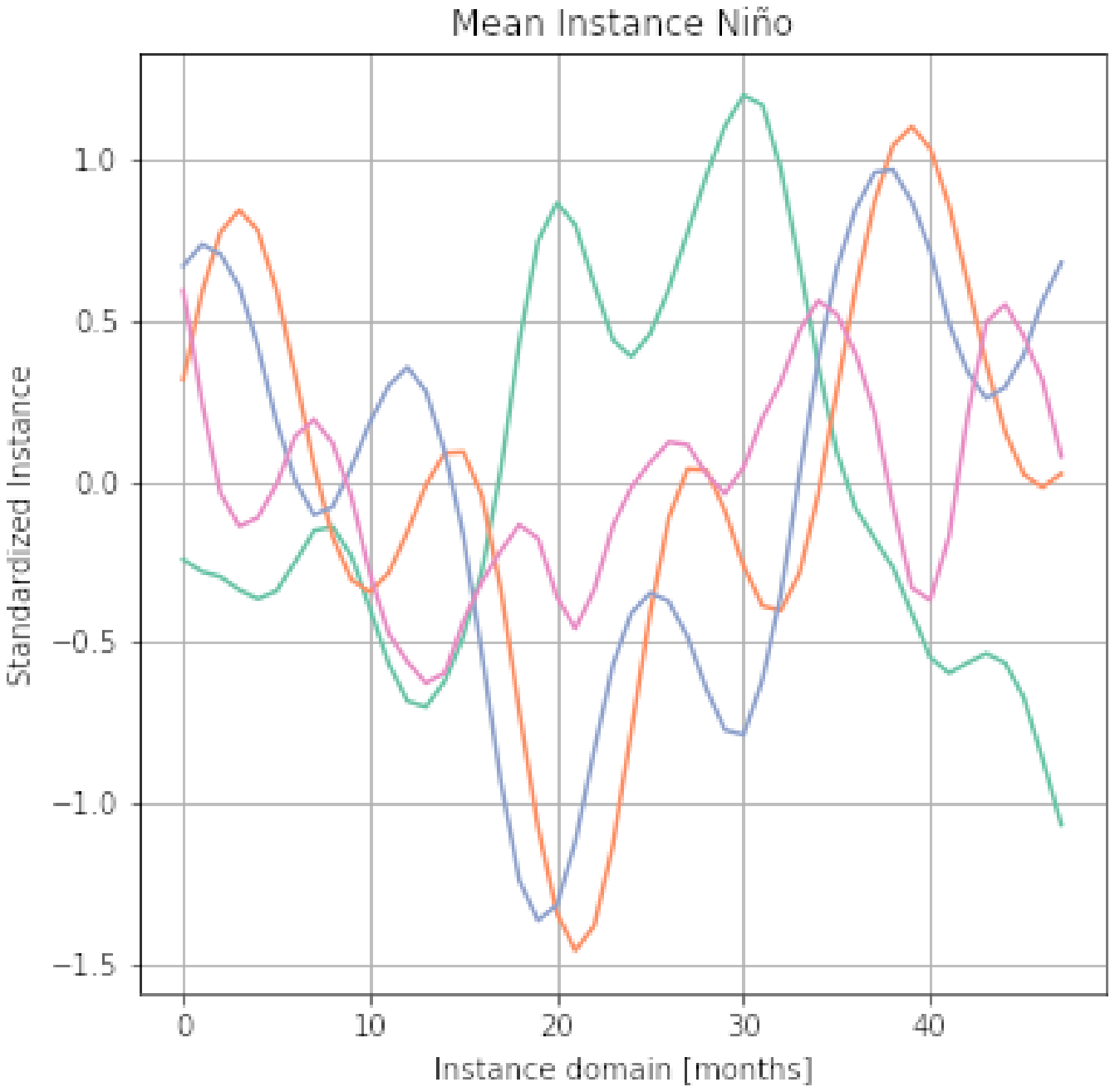}\\
    \caption{}
   \end{subfigure} 
  \begin{subfigure}{0.5\linewidth}
      \centering
    \includegraphics[scale= .35]{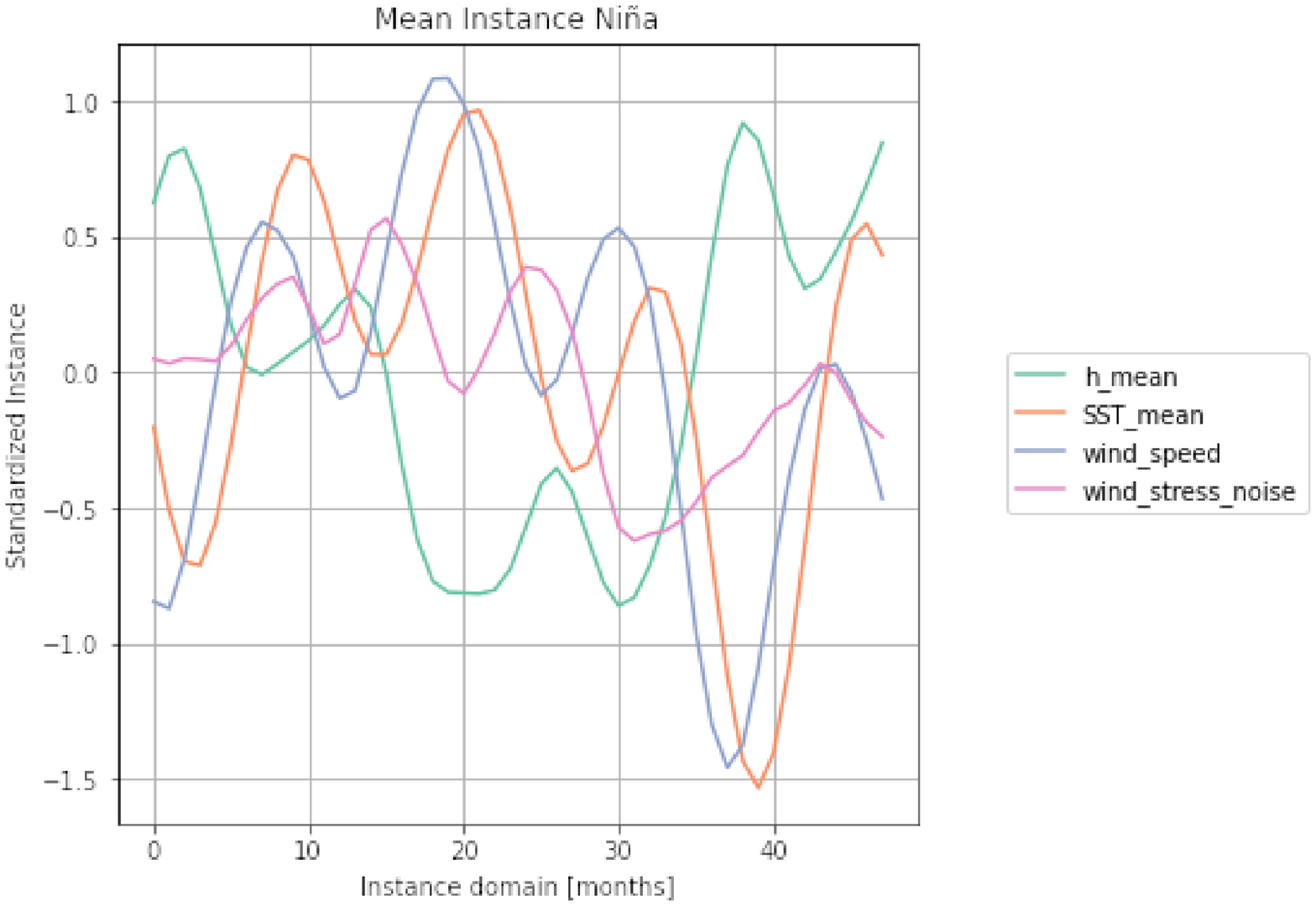}\\
    \caption{}
  \end{subfigure} 
  \begin{subfigure}{0.5\linewidth}
      \centering
    \includegraphics[scale= .35]{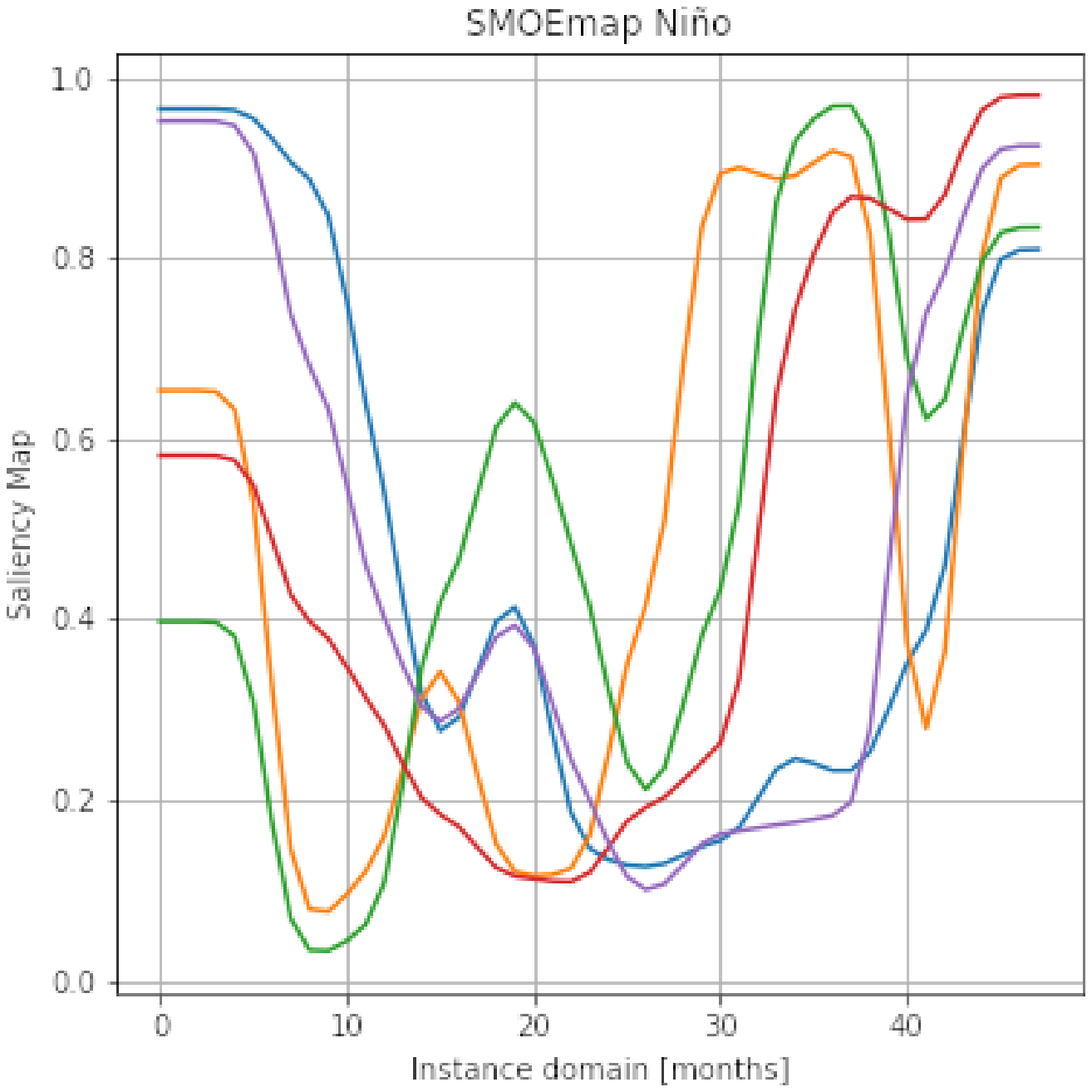} \\
    \caption{}
  \end{subfigure} 
  \begin{subfigure}{0.5\linewidth}
      \centering
    \includegraphics[scale= .35]{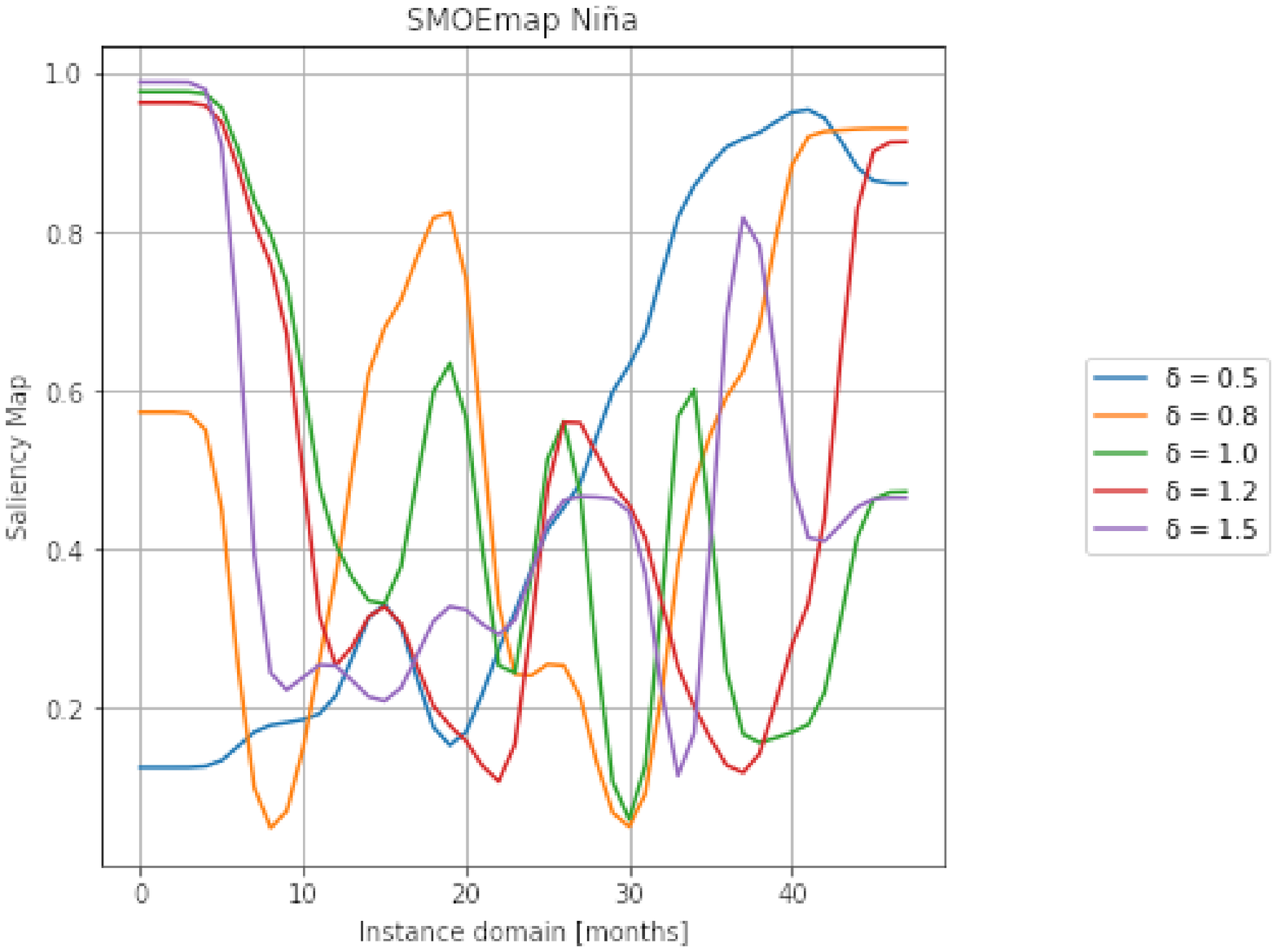} \\
    \caption{}
  \end{subfigure} 
  \begin{subfigure}{0.5\linewidth}
      \centering
    \includegraphics[scale= .35]{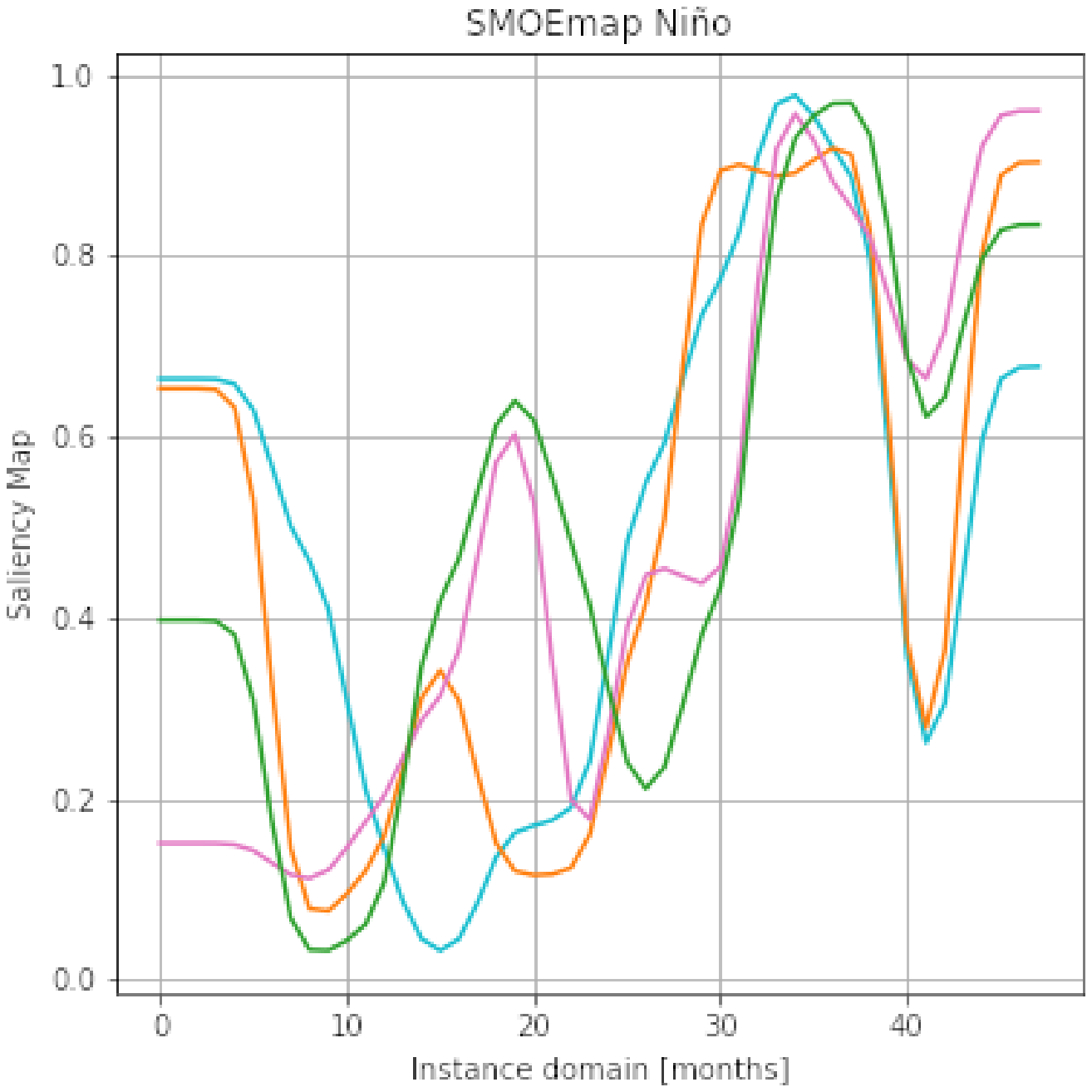} \\
    \caption{}
  \end{subfigure} 
  \begin{subfigure}{0.5\linewidth}
      \centering
    \includegraphics[scale= .35]{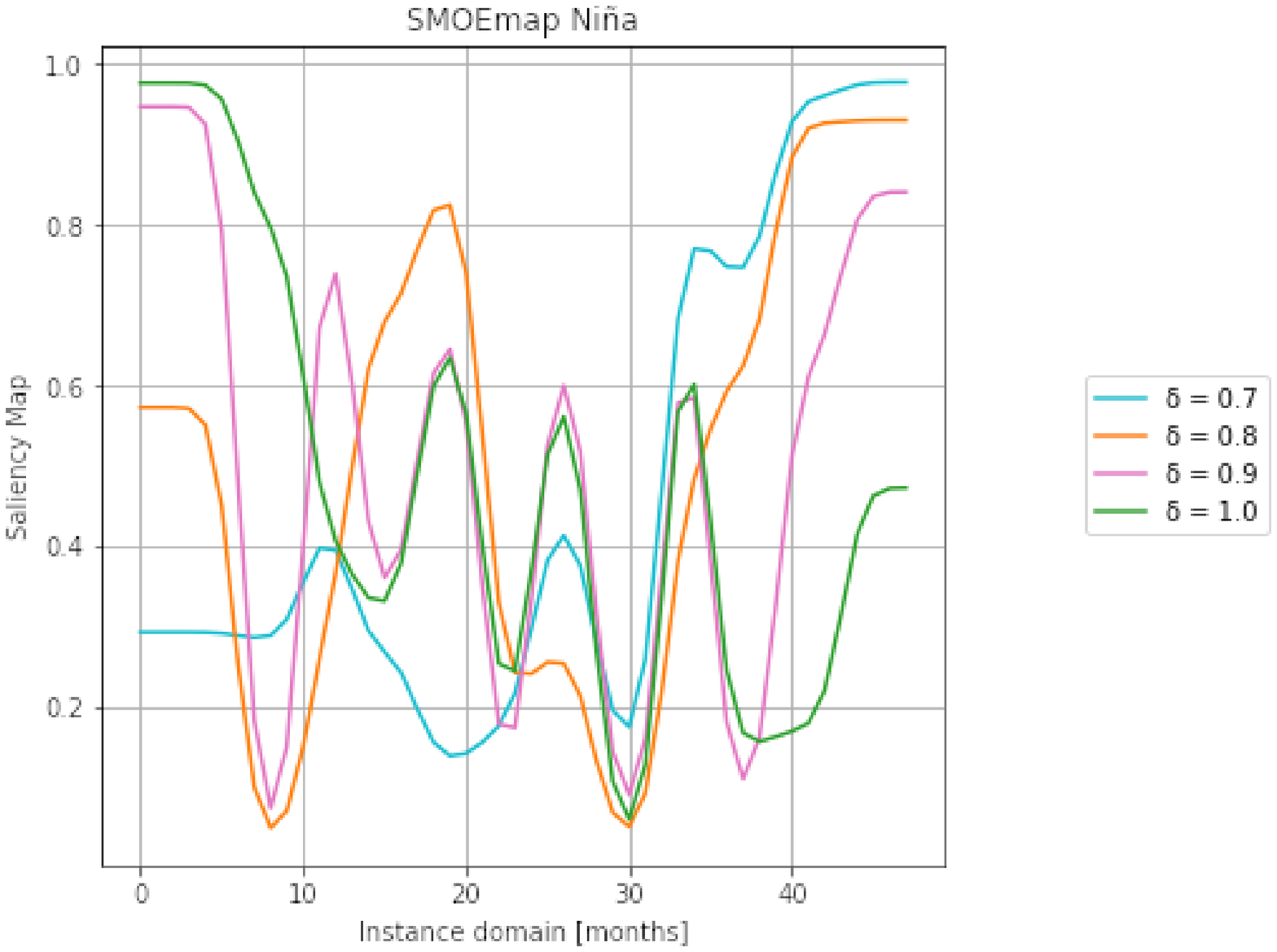} \\
   \caption{}
  \end{subfigure} 
    \caption{Representation of the mean instance of the test data (reference case; Figs. (a)-(b)) and its saliency maps (Figs. (c)-(f)) for the wave distorted case (variation 
    of $\delta$).  
    For the event class El Niño the mean instance and its saliency maps are reported on the left column, while on the right column the results for the class La Niña are shown.}
    \label{fig:smoe_example_delta_sc}
\end{figure}

The application of combined SMOE Scale to the mean instances (namely, the instances obtained by averaging all samples of the test data) can help identifying which patterns in the data are captured by the CNN to generate (accurate or degraded) ONI predictions. 
The reasons of analyzing the mean instance is that it represents the main patterns in the feature time series; the interpretation of the saliency maps of all instances would turn out to be really unpractical. 
The mean instances for both El Ni\~no and La Niña events for all four features are shown in Fig.~\ref{fig:smoe_example_delta_sc}a-b.
The saliency map of the $\delta = 1$ reference case (green curve in Fig.~\ref{fig:smoe_example_delta_sc}c) shows two peaks with intensity 0.6 and 0.8 around month 18 and 35, respectively (note that the instances are 48 months long and that the lead time is 9 months). 
These two regions are the most salient along the whole instance domain for El Ni\~no prediction and are associated with a doublet of peaks in the thermocline depth (Fig.~\ref{fig:smoe_example_delta_sc}a) 
and a doublet of troughs  of both sea surface temperature (Fig.~\ref{fig:smoe_example_delta_sc}c) and wind 
speed (Fig.~\ref{fig:smoe_example_delta_sc}e).  
Therefore, this combination of patterns in the time series features
represents the characteristic that mostly define the El Niño events. 

Likewise, the saliency maps of the mean instance of the event class La Niña achieves values larger than 0.8 at similar temporal locations corresponding to peaks in the thermocline depth and some troughs in the sea surface temperature and wind speed. 
For the cases $\delta = 0.5$ and $\delta= 0.8$ (where waves are propagating faster than the reference case), 
the detection of peaks in the thermocline is associated to troughs in the sea surface temperature and vice versa; this combination of patterns can still lead to a relevant contribution to classify  the ENSO events. 
For the event El Ni\~no, case $\delta = 0.8$ shows a broad and flat peak with intensity larger than 0.8 in the saliency map around month 35 and the CNN is still able to capture patterns located nearly at the same months as  for the reference case; case $\delta = 0.5$ shows high intensities in the saliency map in the region 0-10 months. Here the thermocline depth shows a soft minimum value opposed to the high valued peak in the the sea surface temperature and wind speed. 
A deeper insight into the region $\delta = 0.7-1.0$, where the AUC attaines the higest values, can reveal that the CNN models can capture the similar patterns likewise the case $\delta = 0.8$ (Figs. ~\ref{fig:smoe_example_delta_sc}e-f). 
Indeed, the saliency maps of both cases $\delta = 0.7$ and $\delta = 0.9$ reveal the presence of a  large and broad peak with intensity larger than 0.8 at month 35; the case $\delta = 0.9$  shows that the pattern recognition activity is much more similar to the model trained under the reference case, since its saliency map can point out some other details of interest, e.g. the peak with intensity 0.6 at month 18.  

Similar results are obtained for the event La Ni\~na.
On the other hand, the results  for the cases $\delta = 1.2$ and $\delta = 1.5$ (where waves are propagating slower) reveal that the maximum at 18 months disappears in the saliency map, but the main contributions can be found either around month 35 or at 0-10 months.
For case $\delta = 1.2$, the event El Ni\~no is characterized by the combination of a broad peak 
in both the sea surface temperature and the wind speed with a less important contribution (than in the reference 
case) in the thermocline depth located at months 32-48; for case $\delta = 1.5$ we can observe a behaviour similar to case $\delta = 0.5$ in the saliency map.
Instead, the La Ni\~na event for case $\delta = 1.2$ is mainly
characterized by the opposite feature, i.e. a broad peak in the thermocline depth and a trough in the sea surface temperature 
at located at months 0-10; likewise case $\delta = 1.5$ is characterized by the same feature of $\delta = 1.2$ but located at months 32-48.     
\begin{figure}[!] 
\begin{subfigure}[b]{0.5\linewidth}
    \includegraphics[width=1\linewidth]{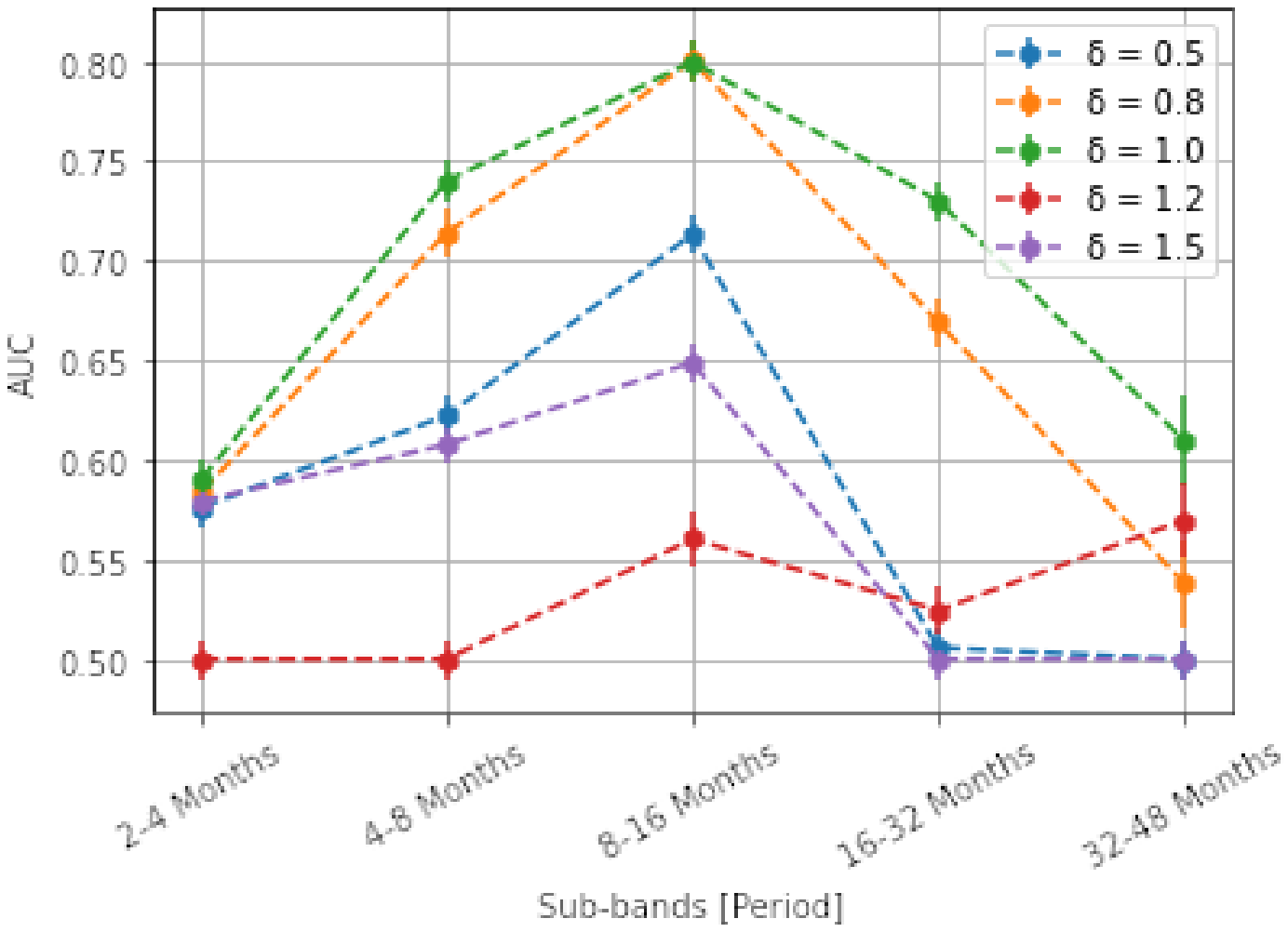}
    \caption{}
  \end{subfigure} 
  \begin{subfigure}[b]{0.5\linewidth}
    \includegraphics[width=1\linewidth]{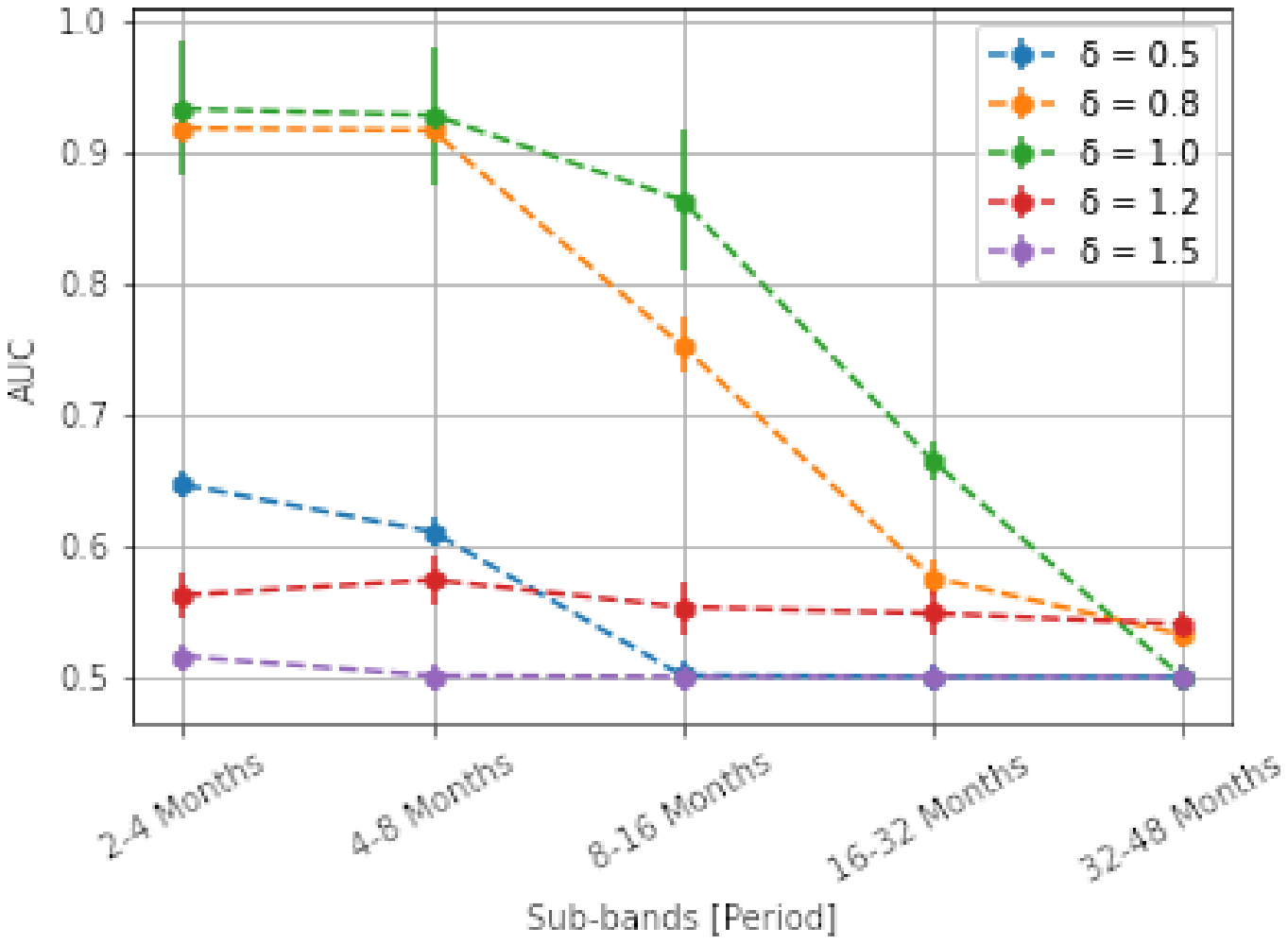}
    \caption{}
  \end{subfigure}
  \caption{The AUC score for different values of $\delta$ for the event El Ni\~no as a function of (a) 
  the band-pass frequency  range  and (b)  the cut-off frequency, obtained by filtering the data by (a) 
  band-pass Butterworth  digital  filter and (b) a low-pass Butterworth digital filter.}
   \label{fig:filering_delta_sc} 
\end{figure}
The application of a band-pass filter on all the sample instances of the test data-set (reference case data) reveals that the information retained in one specific frequency bands can partially retrieve the AUC scores obtained with the non-filtered data. 
The model trained under the reference 
case turns out to be very sensitive to  the frequency bands corresponding to periods 8-16 months with AUC 0.80  (Fig.~\ref{fig:filering_delta_sc}a).
On the contrary, a complete degradation of AUC scores is attained at lower and higher frequency bands, e.g. in both intervals 16-32 months and 2-4 months we observe an AUC value of 0.58. 
Similar results can be found for all other cases taken under consideration, i.e. the band 8-16 months represent the most predictive one with a net degradation of AUC score as soon as slower frequency bands are considered. 
Note that case $\delta = 0.8$ still show some analogies with the reference case; the frequency band 8-16 months is the most predictive with AUC score 0.80.
Therefore, this result points out that the activation of the hidden layers showed with the analysis of the saliency
map can be traced to the detection of oscillating trends with specific carrier frequencies within a low-medium band 
of frequencies. 
The presence of details on a shorter frequency scale (i.e. period 16-32 months), however, is still fundamental and needed to allow the 
CNN to make an accurate classification of the event El Ni\~no.
The smoothing of the sample instances with a low-pass filter (Fig.~\ref{fig:filering_delta_sc}b) reveals the instances 
tend to substantially loose many of their discriminating patterns at cutoff frequencies corresponding to 8 or 16 months.  
For example, in the cases $\delta= 0.8$ and $\delta= 1.0$ we can observe a decrease of the predictive power with a 
degradation of 0.1 AUC at 8 months and 0.3 AUC at 16 months.
Hence, medium-low frequency patterns (4-8 months) as those contained in 
the thermocline depth or in the wind-noise time series can play an important role in the detection of the events. 
\begin{figure}[!] 
\begin{subfigure}{0.4\linewidth}
    \includegraphics[width=1\linewidth]{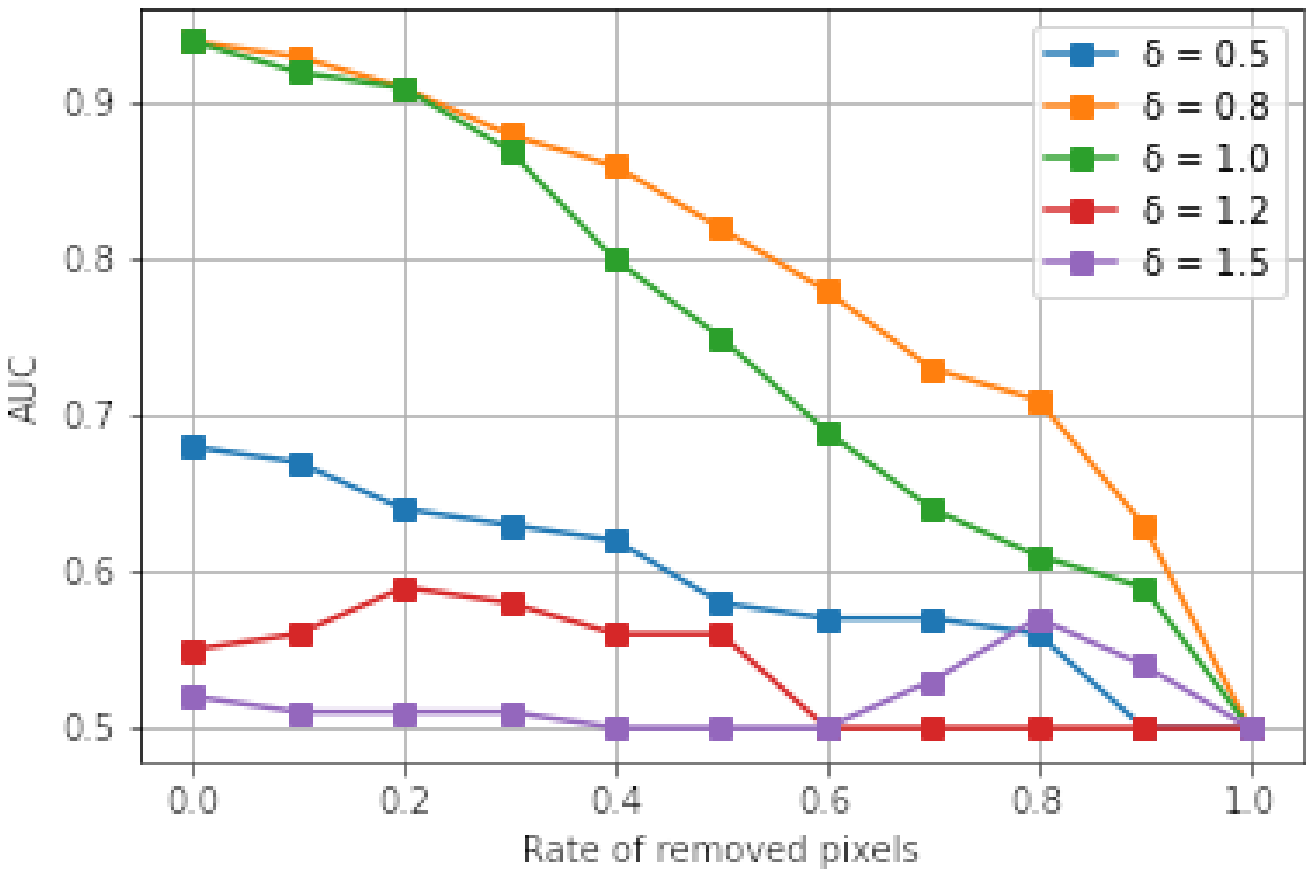}
    \caption{}
  \end{subfigure} 
  \begin{subfigure}{0.4\linewidth}
    \includegraphics[width=1\linewidth]{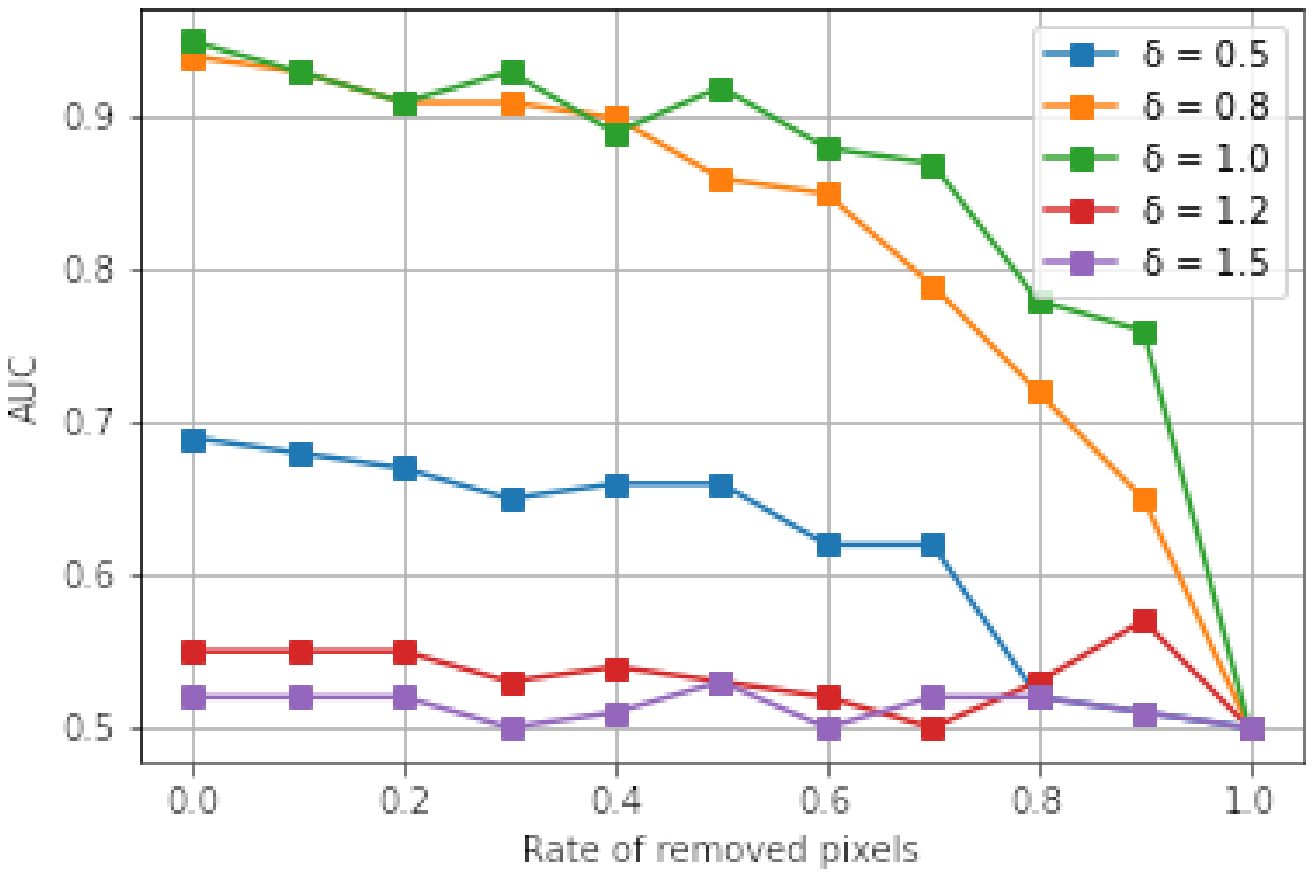}
    \caption{}
  \end{subfigure} 
  \caption{Evaluation of AUC when the ROAR method (a) or the replacing at random strategy is applied (b); on the x-axis the ratio of pixels replaced and on the y-axis the AUC value.}
   \label{fig:ROARs} 
\end{figure}
Finally, to ensure the correct implementation of the combined SMOE Scale and to guarantee the validity of the results obtained, we used (and adapted to our analysis) the metrics ROAR (Remove and Retain) introduced in \cite{mundhenk2019efficient}.
The replacement in the validation sets of an increasingly amount of salient spatial locations with zero-valued pixels rapidly deteriorates the predictive characteristics of the data.
It is worthwhile to mention that the CNN does not train any bias term neither in the convolutional layers nor in the 
dense layers. Thus zero-valued patterns are considered as absolutely non-informative, i.e. the propagation of a zero-valued pattern through the CNN is designed to prevent the activation of any stimulus along the hidden layers.
In Fig.~\ref{fig:ROARs}a we can observe that the removal of the top 50\% of the salient pixels via ROAR (actually 24) guarantees a considerable decrease of the AUC. 
In fact, under the reference case model the AUC scores present a loss equal to 0.20. 
Contrary to this, when randomly replacing the 50\% pixels with zero-valued pixels we can still observe a slighter decrease in the 
AUC curve under the reference, i.e. a loss equal to 0.03.
In addition, similar results can be found when even 
considering all the distorted physics data analyzed above (Fig.~\ref{fig:ROARs}b).

\subsection{ Upwelling feedback} 
\begin{figure}[!]
    
    \includegraphics[width=\linewidth]{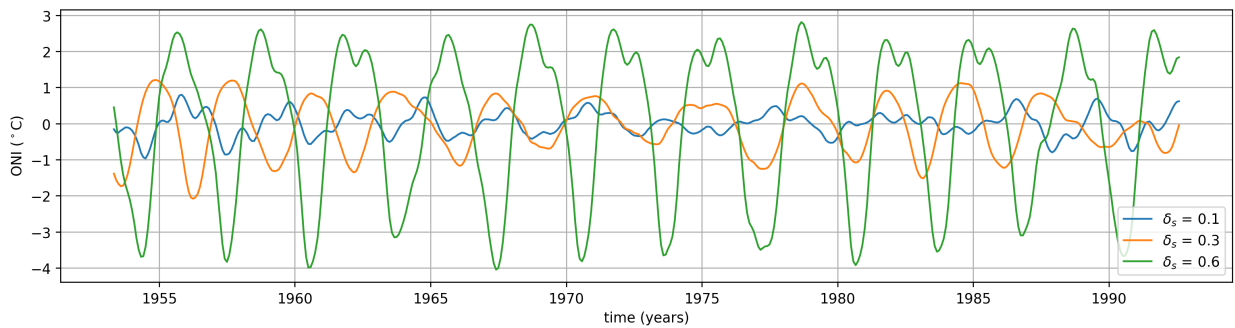}
    \caption{Several time series of ONI calculated from ZC model simulations using $\delta_s$ parameter 
    values of 0.1, 0.3 and 0.6 using $\mu = 2.7$. }
    \label{fig:ONI_distorted_delta_s}
\end{figure}

We next consider the distortion of the model data due to a wrong representation of the upwelling feedback, represented
by the parameter $\delta_s$ in the ZC model. 
Figure \ref{fig:ONI_distorted_delta_s} shows that the ONI's amplitude increases (decreases) for larger (smaller) 
values of $\delta_s$. This behavior is expected because the upwelling feedback is a positive one, enhancing the 
existing sea surface temperature anomaly further and consequently increasing  the amplitude of the ONI. 
\begin{figure}[!] 
\begin{subfigure}{0.5\linewidth}
    \includegraphics[width=1\linewidth]{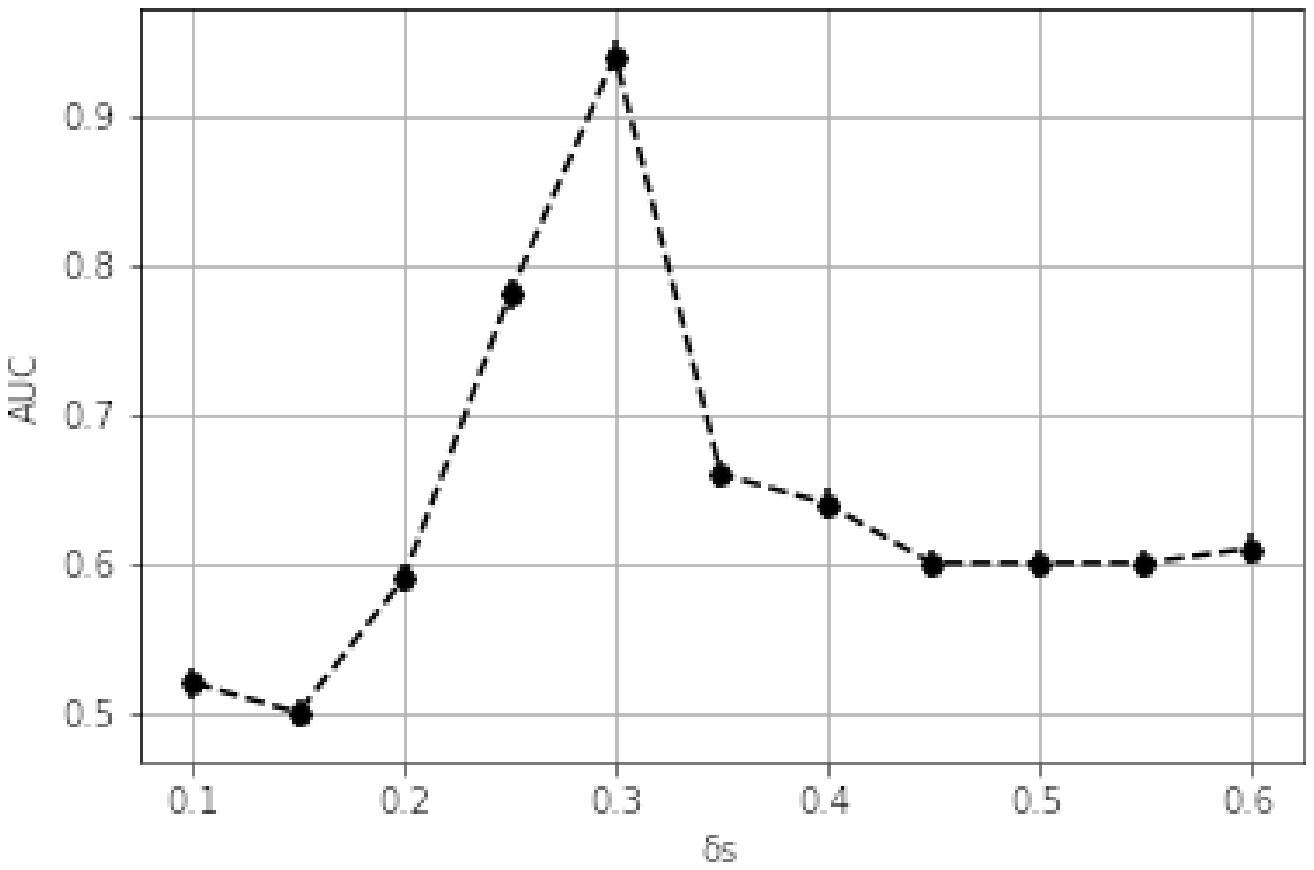} 
    \caption{}
  \end{subfigure} 
  \begin{subfigure}{0.5\linewidth}
    \includegraphics[width=1\linewidth]{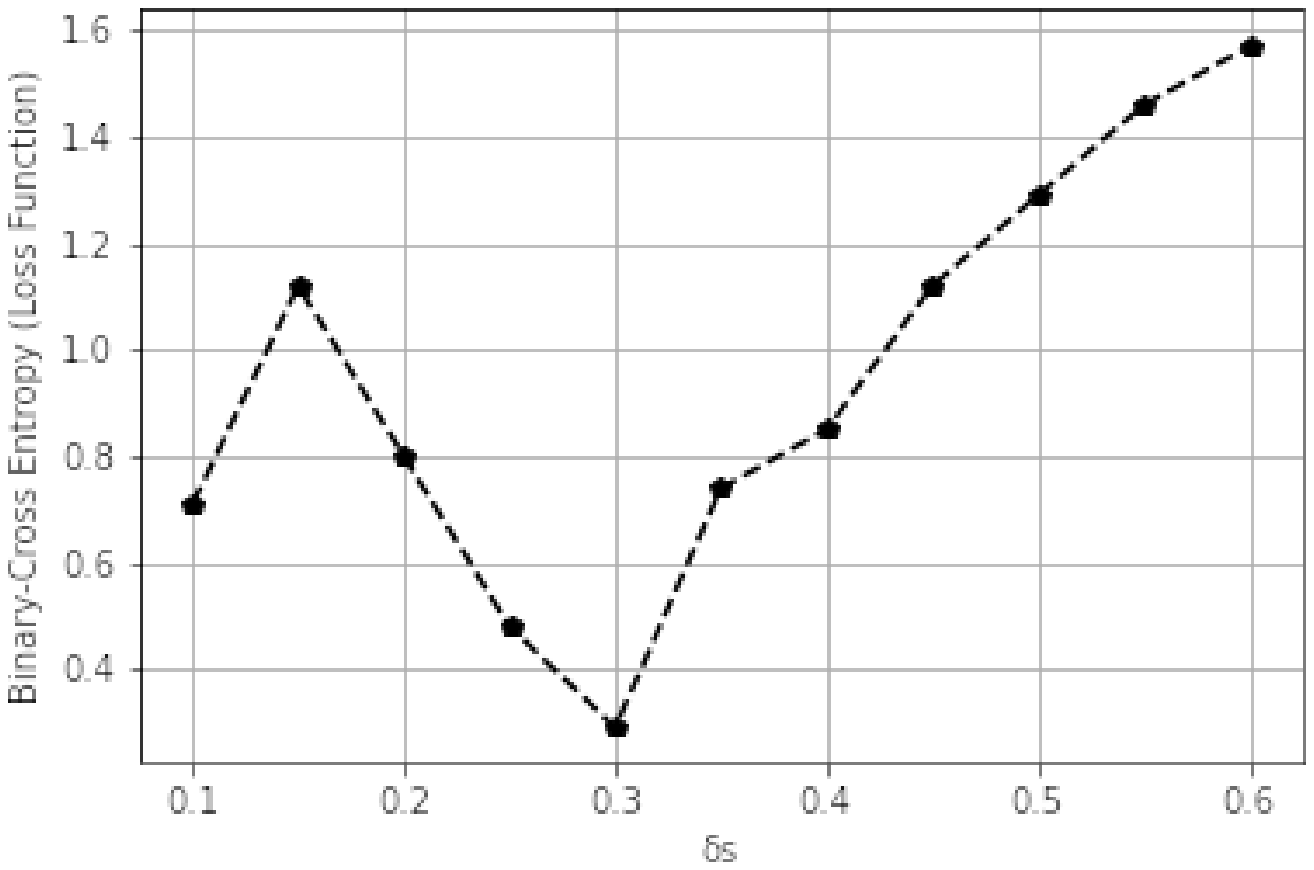} 
    \caption{}
  \end{subfigure} 
  \caption{The AUC score (a) and the loss function (b) as a function of the upwelling feedback parameter $\delta_s$. Each point represents the mean AUC over 5 different folds; error bars are evaluated via standard error mean.}
   \label{fig:auc_deltas} 
\end{figure}
The AUC score versus $\delta_s$ curve (Fig.~\ref{fig:auc_deltas}a)  reveals that a particular tuning of the parameter 
$\delta_s$ strongly affects the accuracy of the CNN model when trained with distorted data. 
By construction, the AUC score attains the highest score  at the reference value $\delta_s= 0.3$ (AUC 0.94).  
For $\delta_s < 0.3$ the profile of the curve suggests a net degradation in the AUC scores with the lowest score 
attained at $\delta_s = 0.15$ (AUC 0.5), whereas at $\delta_s>0.3$ the AUC scores remain stable, but still 
attaining values lower than 0.7.  The profile of the UAC has a plateau at values of 0.6 as $\delta_s$ goes 
towards the boundary value $\delta_s = 0.6$. The evaluation of the loss function  (Fig.~\ref{fig:auc_deltas}b)  
as a function of the parameter $\delta_s$ 
confirms the results obtained above. At $\delta_s = 0.3$ the global minimum is achieved, the net degradation 
occurring at lower and higher $\delta_s = 0.3$ are still present; the loss function increases 
monotonically in both cases.  
\begin{figure}[!]
    \begin{subfigure}{0.5\linewidth}
    \centering
    \includegraphics[scale= 0.33]{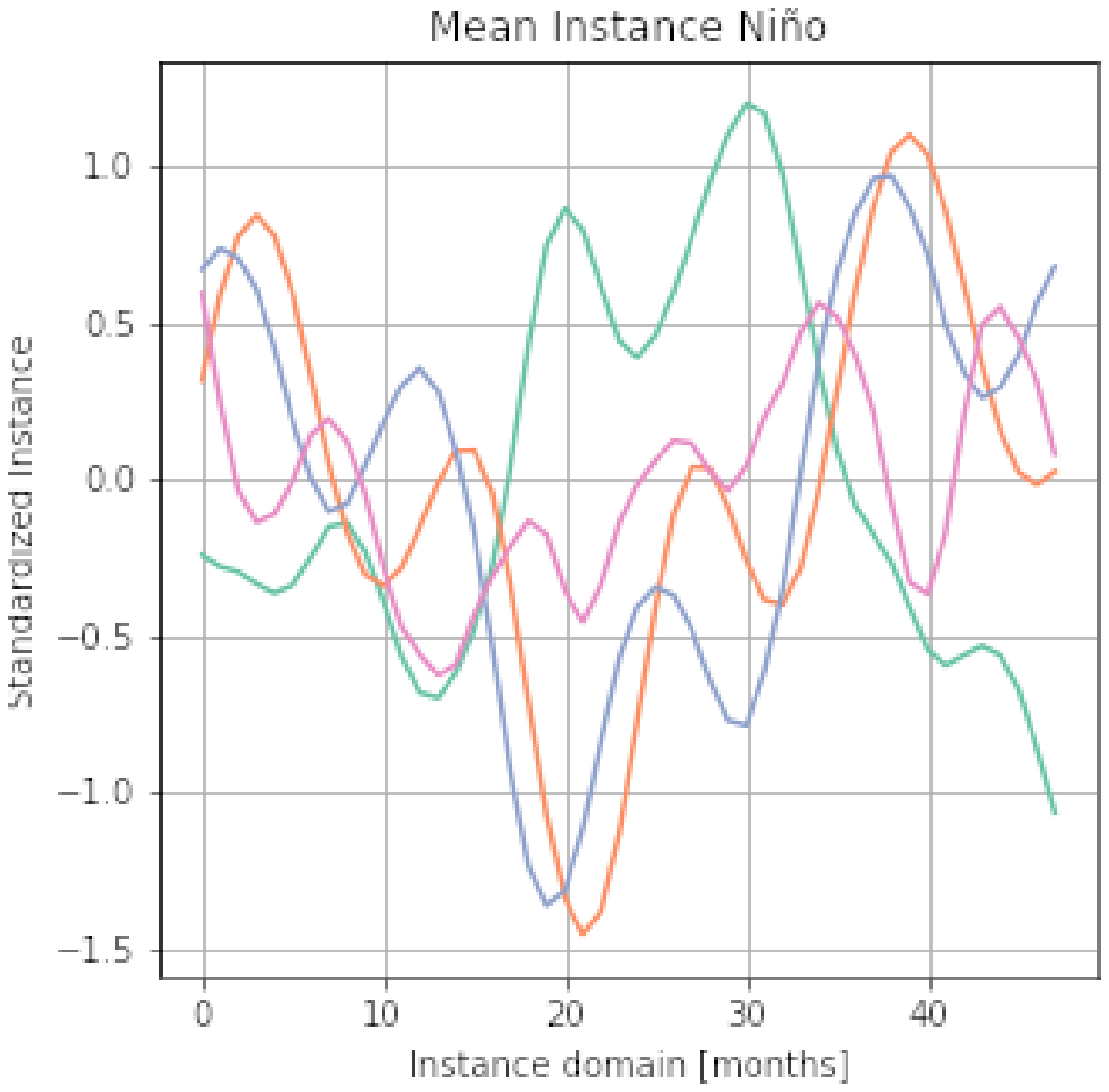}\\
    \caption{}
   \end{subfigure} 
  \begin{subfigure}{0.5\linewidth}
     \centering
    \includegraphics[scale= 0.33]{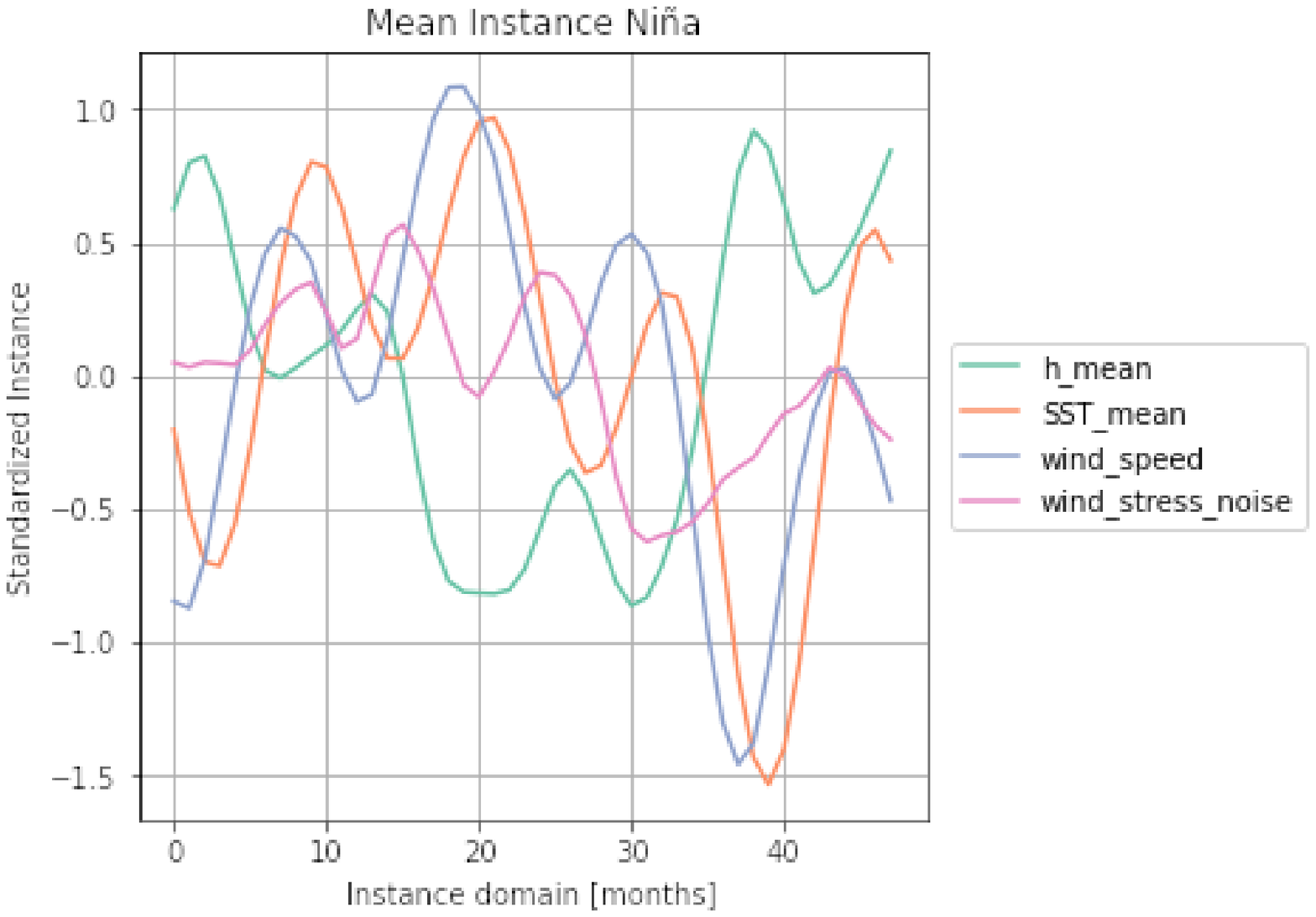}\\
    \caption{}
  \end{subfigure} 
  \begin{subfigure}{0.5\linewidth}
  \centering
    \includegraphics[scale= 0.33]{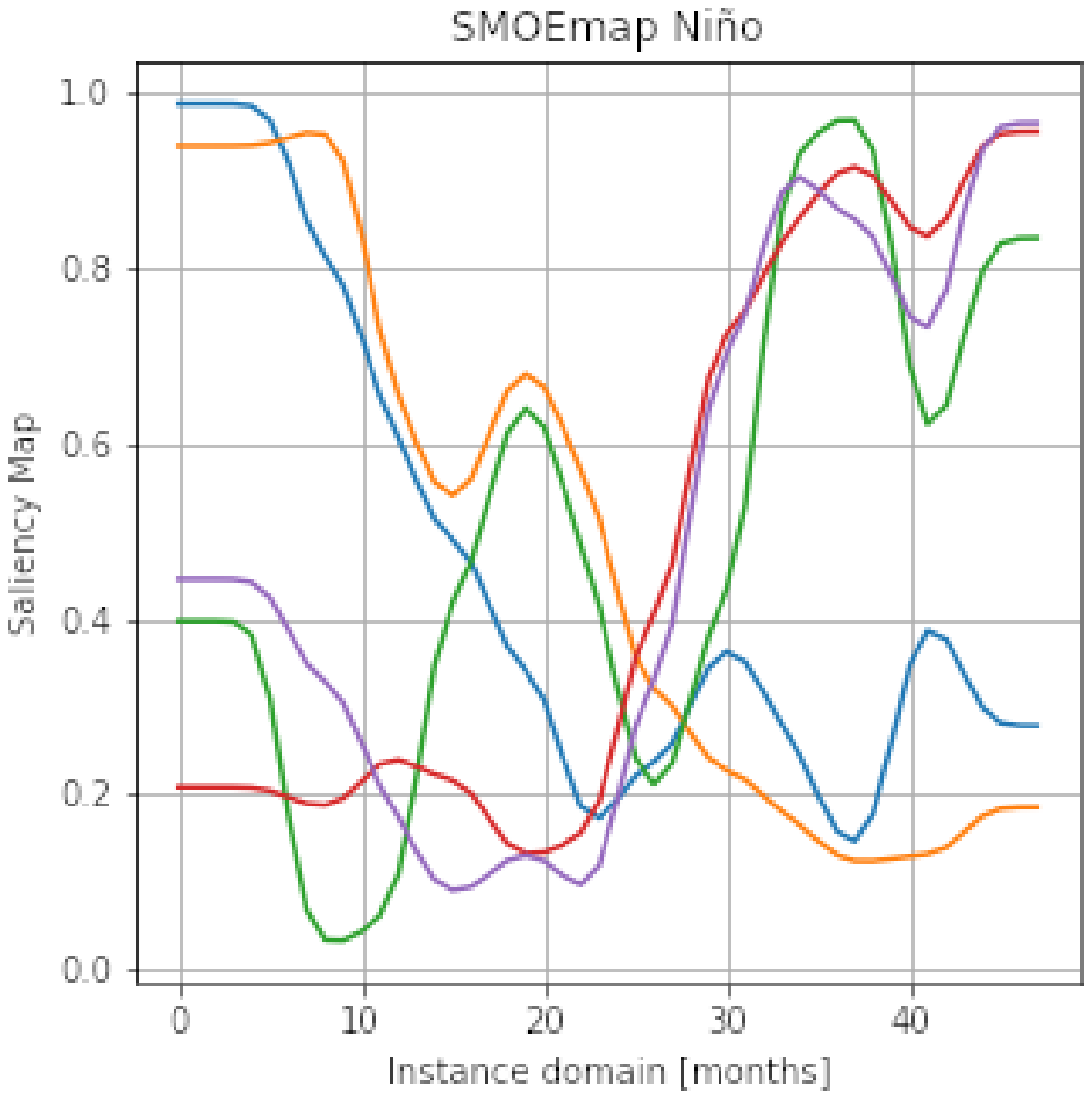} \\
    \caption{}
  \end{subfigure} 
  \begin{subfigure}{0.5\linewidth}
  \centering
    \includegraphics[scale= 0.33]{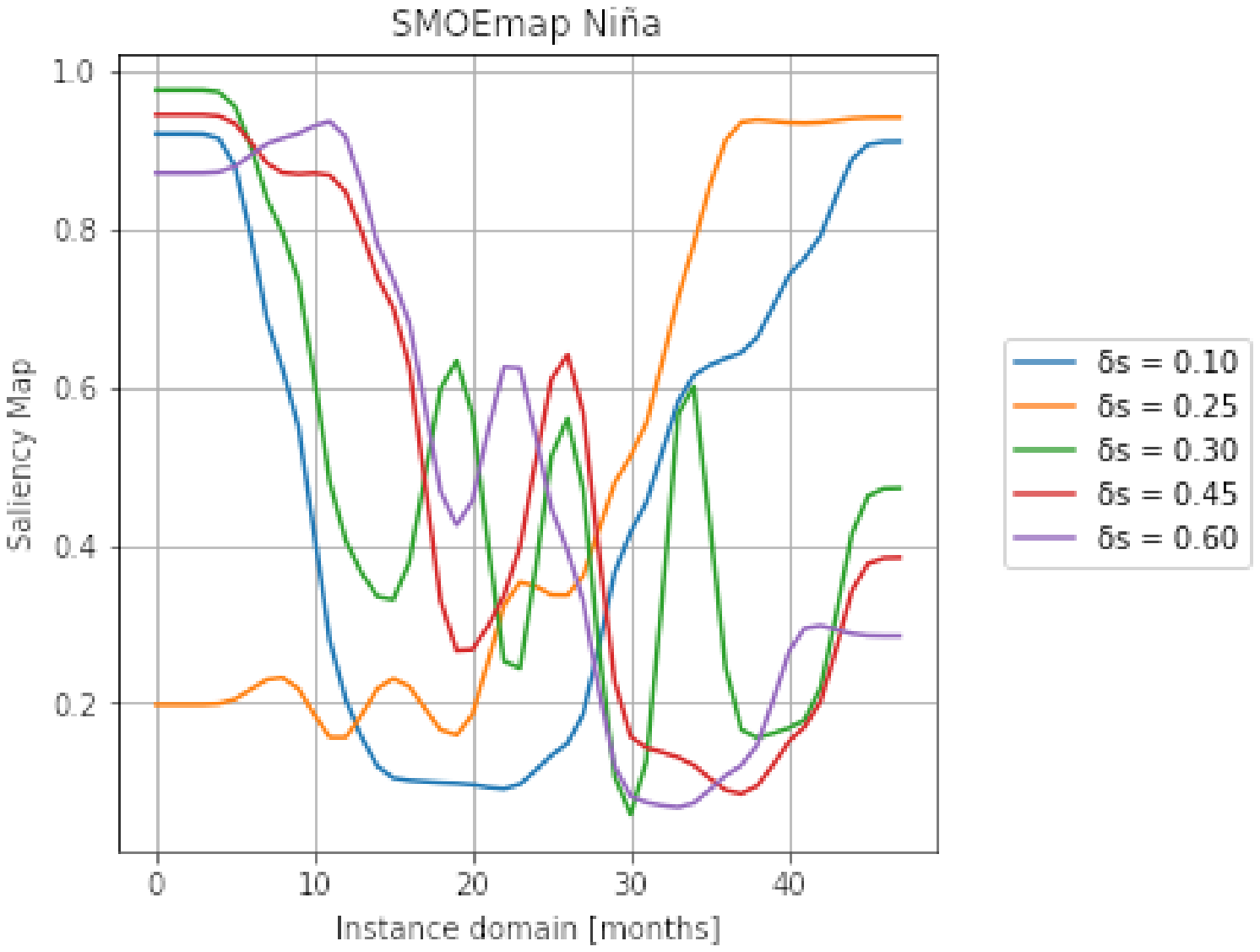} \\
    \caption{}
  \end{subfigure} 
    \caption{Representation of the mean instance of the test data (refrence case; Figs. (a)-(b)) and its saliency maps (Figs. (c)-(d)) for the upwelling distorted case(variation of $\delta_s$).  
    The left column contains the mean instance and the saliency map for the event class El Niño,  while on the right column the mean instance and the saliency map for the class La Niña are shown.}
    \label{fig:smoe_example_deltas_sc}
\end{figure}

Similarly to the analysis provided for the distortion of the $\delta$ parameter, we nextconsider the mean instance (of the reference case data) and its the saliency maps.  
For El Ni\~no events, we can observe that different regions of saliency can be associated to different variation of $\delta_s$, i.e. for $\delta_s < 0.3$ the saliency map indicate the left part of the instance as the most predictive and  for $\delta_s > 0.3$ the right part; for El Ni\~na events the opposite  is valid.
In particular, cases $\delta_s = 0.10$ and $\delta_s = 0.25$ turn out to be very salient around 0-10 month in the saliency map with intensity above 0.8 where a peak occur in both sea surface temperature and wind speed time series. The case $\delta_s = 0.25$ has actually a larger salient region in the saliency map up to month 20 and includes more interesting patterns, e.g. a more complete sequence of oscillations and a deep trough in the sea surface temperature time-series feature.
On the contrary, for cases $\delta_s = 0.45$ and $\delta_s = 0.60$ the saliency maps achieve intensities larger than 0.8 around 32-48 months and capture one single broad oscillating peak in the sea surface temperature time-series feature.
For La Ni\~na event, we observe that saliency maps of cases $\delta_s = 0.10$ and $\delta_s = 0.25$ present intensities higher than 0.8 at 42-48 months. 
It is interesting to observe that the saliency map of case $\delta_s = 0.25$ goes above intensity 0.8 and presents a plateau around 32-48 months. Similarly to the El Ni\~no events a deep trough in the sea surface temperature time series feature is captured by the CNN.
\begin{figure}[!] 
\begin{subfigure}[b]{0.4\linewidth}
    \includegraphics[width=1\linewidth]{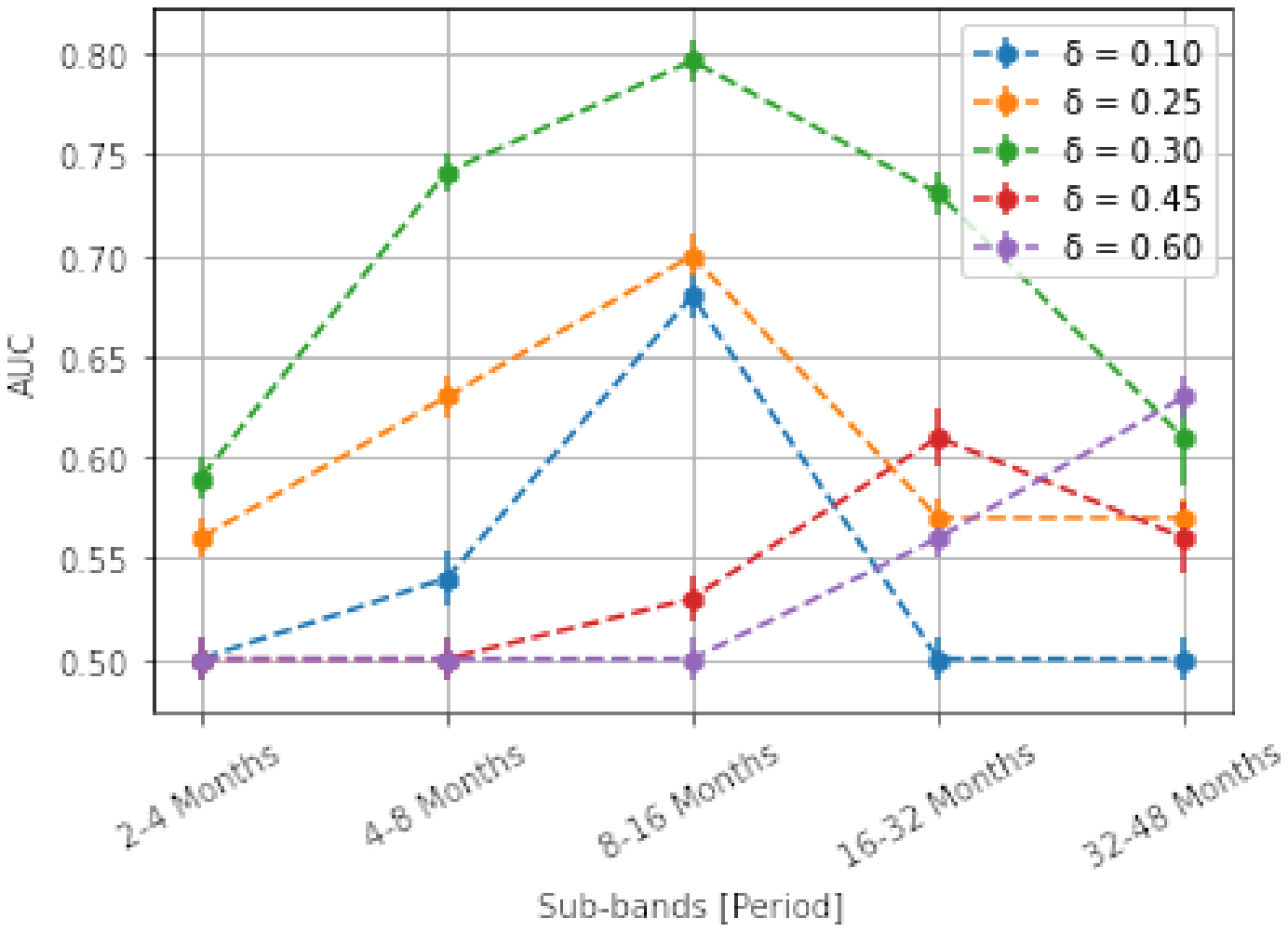}
  \end{subfigure} 
  \begin{subfigure}[b]{0.4\linewidth}
    \includegraphics[width=1\linewidth]{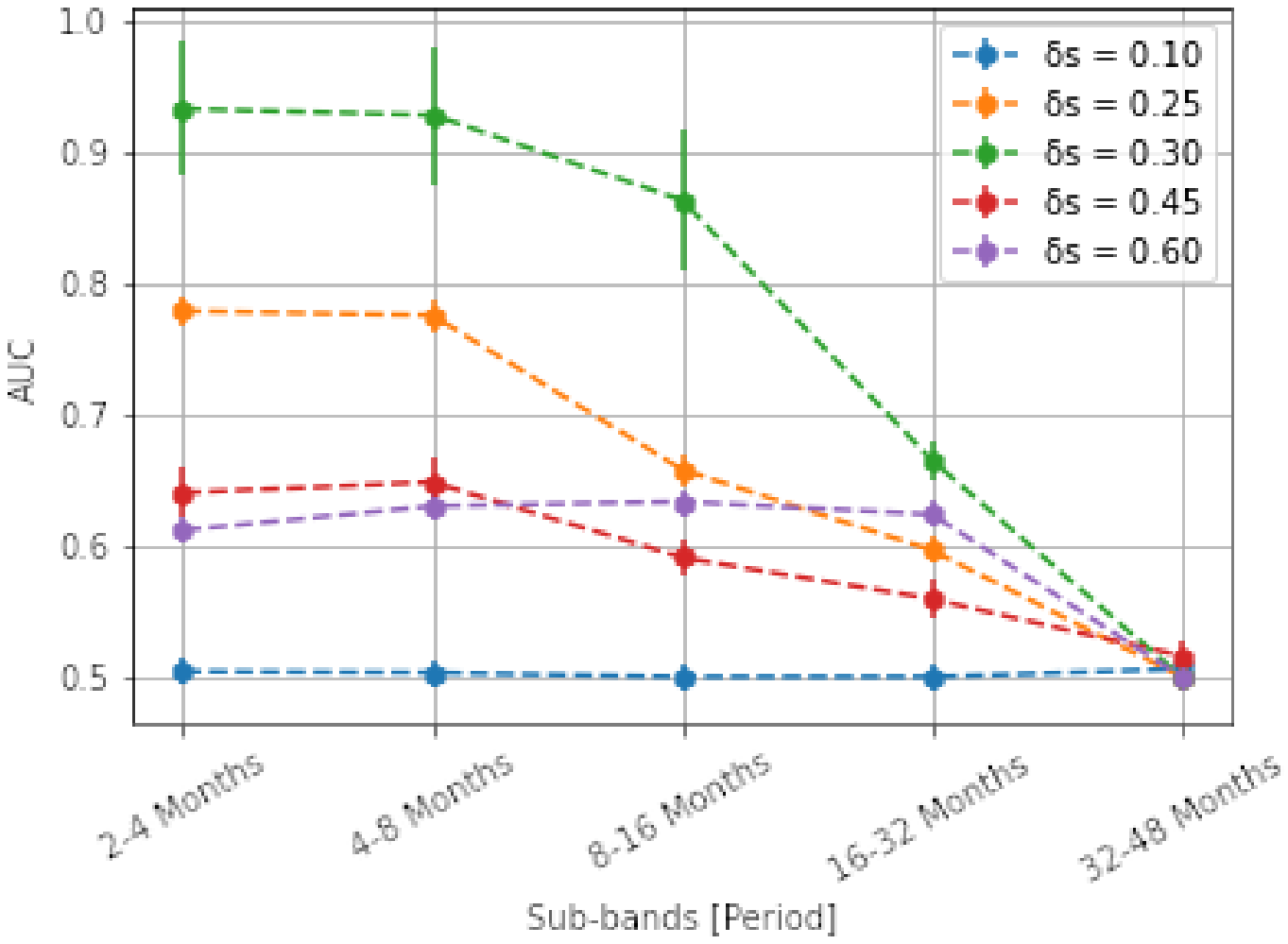}
  \end{subfigure}
  \caption{The AUC score for different values of $\delta_s$ for the event El Ni\~no as a function of a) the band-pass frequency 
  range  and b)  the cut-off frequency, obtained by filtering the data by a) band-pass Butterworth 
  digital  filter and b) a low-pass Butterworth digital filter. }
   \label{fig:filtering_deltas_sc} 
\end{figure}

The application of band-pass and low-pass filters on the sample instances bring to light a result similar to the analysis done for the parameter $\delta$.
As we can see in Fig.~\ref{fig:filtering_deltas_sc}a, the 
instances for the case $\delta_s = 0.25$ can partially retrieve the original prediction with AUC 0.70 when a band-pass filter with bandwidth 8-16 months is applied, whereas for the other cases such 
as $\delta_s = 0.6$ the original predictions can be retrieved only by oscillations lying within the frequency band corresponding to 32-48 months. 
The smoothing of the instances via low-pass filter
(Fig.~\ref{fig:filtering_deltas_sc}b)  shows  that the removal of high-frequency patterns oversimplifies 
the data; and so the classification task cannot be solved relying on the information contained in the low-frequency data only. 

As confirmed by the filtering of the instances, the frequency bands 4-8 months and 8-16 months represent 
the main frequency bands in the reference case ($\delta_s= 0.3$). Capturing one of these two can retrieve a 
considerable amount of skill. The case $\delta_s= 0.25$ focuses a large amount 
of relevant patters mainly in the frequency band 8-16 months. The filtering with a low-pass digital filters also 
reveals that a cut-off frequency of 16 months can reduce the AUC in both cases, but a cut-off frequency
of  8 months leads to a degradation for the reference case only. In the latter scenario, we register a loss of 0.1 
AUC, i.e. a degradation on the same order of magnitude as when testing the reference case data and the 
data of case $\delta_s= 0.25$. Hence, this example shows how a manipulation in the intrinsic characteristic 
of the instances can lead to a reduction and oversimplification of the instances, i.e.  the distortion of the 
periodicity of data provokes a reduction or missing of some patterns that are fundamental in the classification 
of the reference case data.   

\subsection{Comparison of CNN and GDNN}

To provide a comparison, we also applied the distorted physics approach in the Gaussian Density 
Neural Network (GDNN) as used in \cite[]{Petersik2020}. 
The Gaussian density terminology refers to the networks purpose of predicting a Gaussian distribution by producing
both a mean and standard deviation as output.  The variable to be predicted (or target variable) is also the ONI at a 
(lead) time in the future. The features used in the GDNN are described by \cite{Petersik2020}: ONI, network 
graph connectivity metric $c_2$, adjusted Hamming distance $\mathcal{H}^*$ (measure of change in the network graph) 
and a seasonal cycle (SC) in the form of a cosine. The warm water volume (WWV, volume of water above the 
20 $^\circ$C thermocline) is not available in the output of ZC model and  therefore the thermocline depth 
itself was  used here. All feature datasets are normalized before training.  

Training the GDNN consists of a number of ensemble members that are trained in parallel. Each of the members is 
trained for 100 iterations over 500 epochs with a batch size of 100. The training starts with a random selection of 
hyperparameters within bounds defined by the user and is then optimized using the ADAM  algorithm 
\cite[]{kingma_adam:_2014}  with a user specified learning rate, drop out and Gaussian noise. The resulting 
ensemble members  each predict a mean and standard deviation of the target  variable and  these predictions 
are then averaged over the ensemble for the final prediction. Again the lead time is 9 months in the 
result below. 

\begin{figure}[!] 
  \begin{subfigure}[b]{0.45\linewidth}
      \includegraphics[width=.9\linewidth]{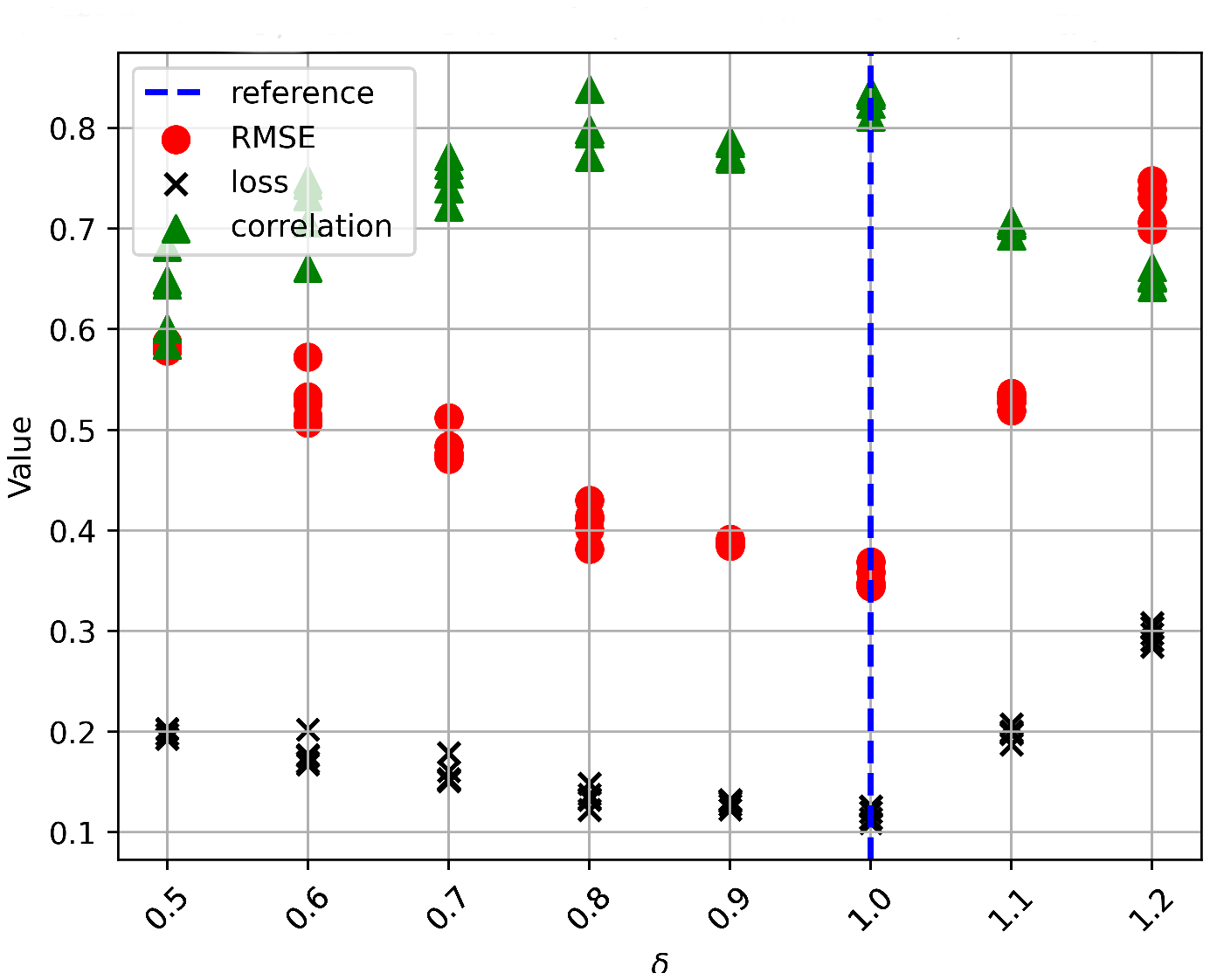} 
    \label{metrics1a}
  \end{subfigure} 
  \begin{subfigure}[b]{0.45\linewidth}
    \includegraphics[width=.9\linewidth]{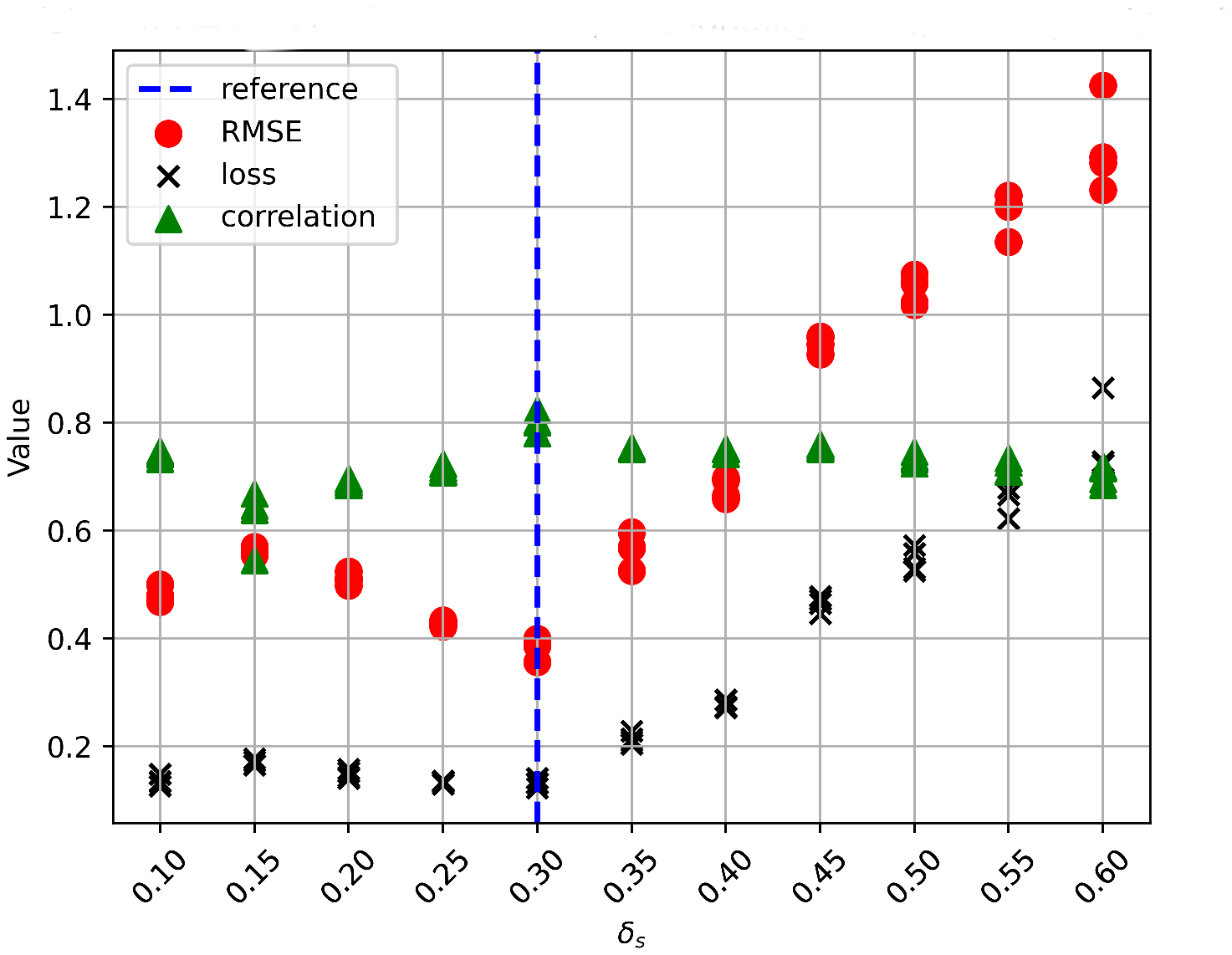}
    \label{metrics1b}
  \end{subfigure}
  \caption{Performance of the GDNN when trained on distorted ZC model data using 
  several values of (a) $\delta_s$ and (b)  $\delta$. 
}
\label{fig:GDNN_performance_delta}
\end{figure}

We use two different measures for the performance of the GDNN: the RMSE
and  the Pearson correlation; also the loss function is shown. Different simulations 
give different networks and  give  different performance values. 
The GDNN's, when trained on distorted physics data,  still perform 
consistently when varying  $\delta$ or  $\delta_s$.  However,  a change 
in the ONI's amplitude in the training data (such as for higher than reference 
$\delta_s$) is  poorly corrected for, leading to a large overestimation of the 
predicted variable (e.g. see $\delta_s = 0.40$ in Fig.~\ref{fig:GDNN_performance_delta}). The model 
only tolerates a difference in amplitude between test and training dataset ONI if only a small distortion of the 
variable is used (e.g. $\delta_s = 0.35$). The ability to compensate for period but not amplitude is  explained 
by the relatively simple architecture of the GDNN. Whereas the former only requires a scalar addition to the 
input,  the latter would require some linear combination  of (co)sines to be learned by the neural network.  


The attempt of comparing the capability of both CNN and GDNN in  detecting El Ni\~no  
events is made complicated by the intrinsic design of both  models.
Although both the models are trained to solve the same problem, we have to take into 
account that the CNN model is a binary classifier, while the GDNN is designed to solve 
regression problems.
In addition, the fact that both models optimize the same loss function does not ensure 
a relation or a similarity about what the two models learn during the training phase can be found.
The two models could focus on capturing totally different features of data, because the 
outputs of the two models  represent two different probabilities, i.e. the CNN estimates the 
probability of the event itself, whereas the GDNN estimates the probability distribution of the 
ONI index.    However, the ENSO events are based on the behavior of the ONI index and 
we can exploit this fact to make the outputs of the GDNN more close to those of the CNN.
After training the GDNN, we can use the estimation on the Gaussian density to estimate 
the probability of El Ni\~no events, i.e. the probability that the absolute value of ONI index 
is greater than 0.5 $^\circ C $. Thereafter, we can use the AUC metric to compare the 
performance of the two models. 
\begin{figure}[!]
\begin{subfigure}{0.45\linewidth}
    \includegraphics[width=1\linewidth]{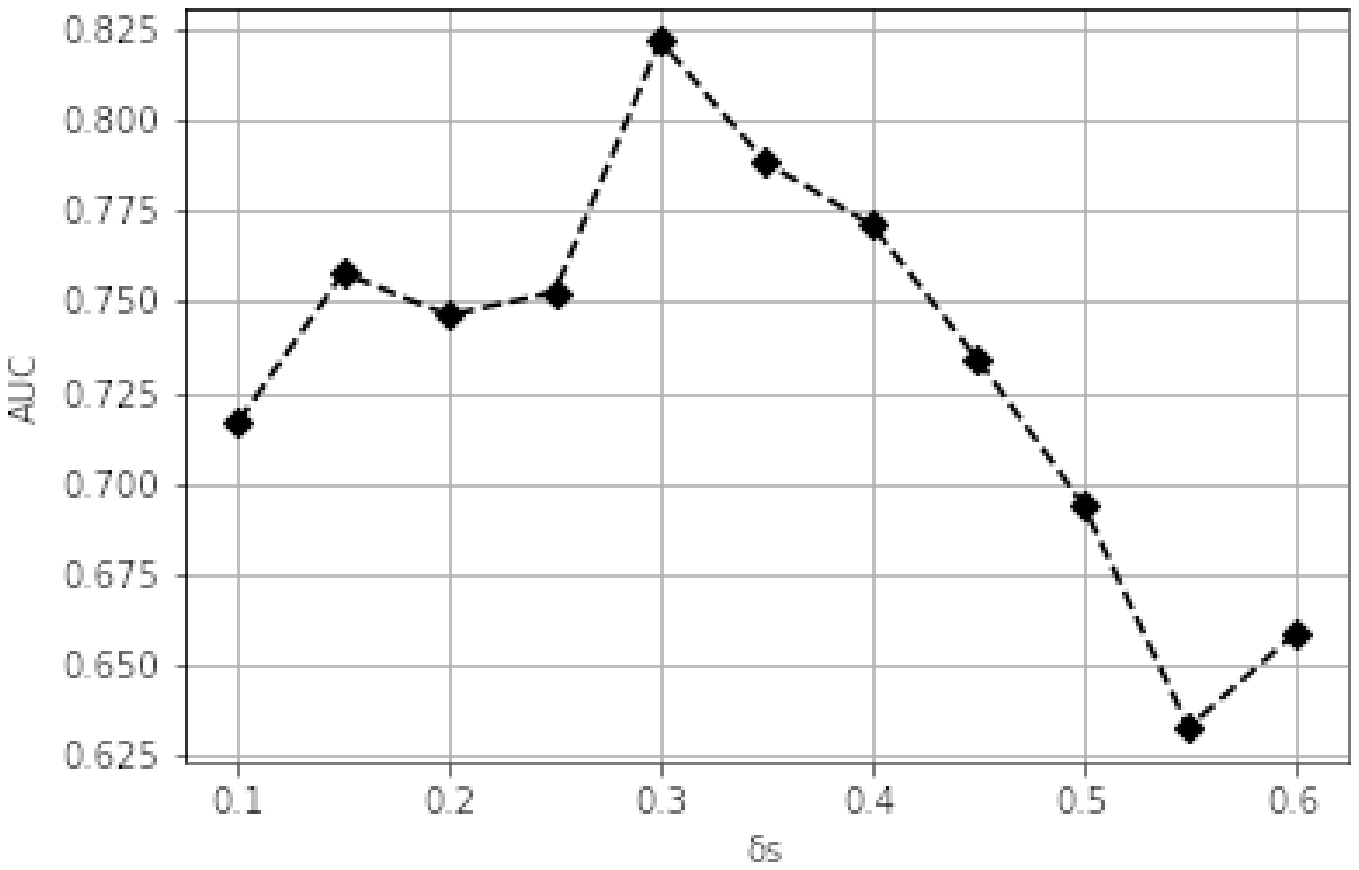} 
  \end{subfigure} 
  \begin{subfigure}{0.45\linewidth}
    \includegraphics[width=1\linewidth]{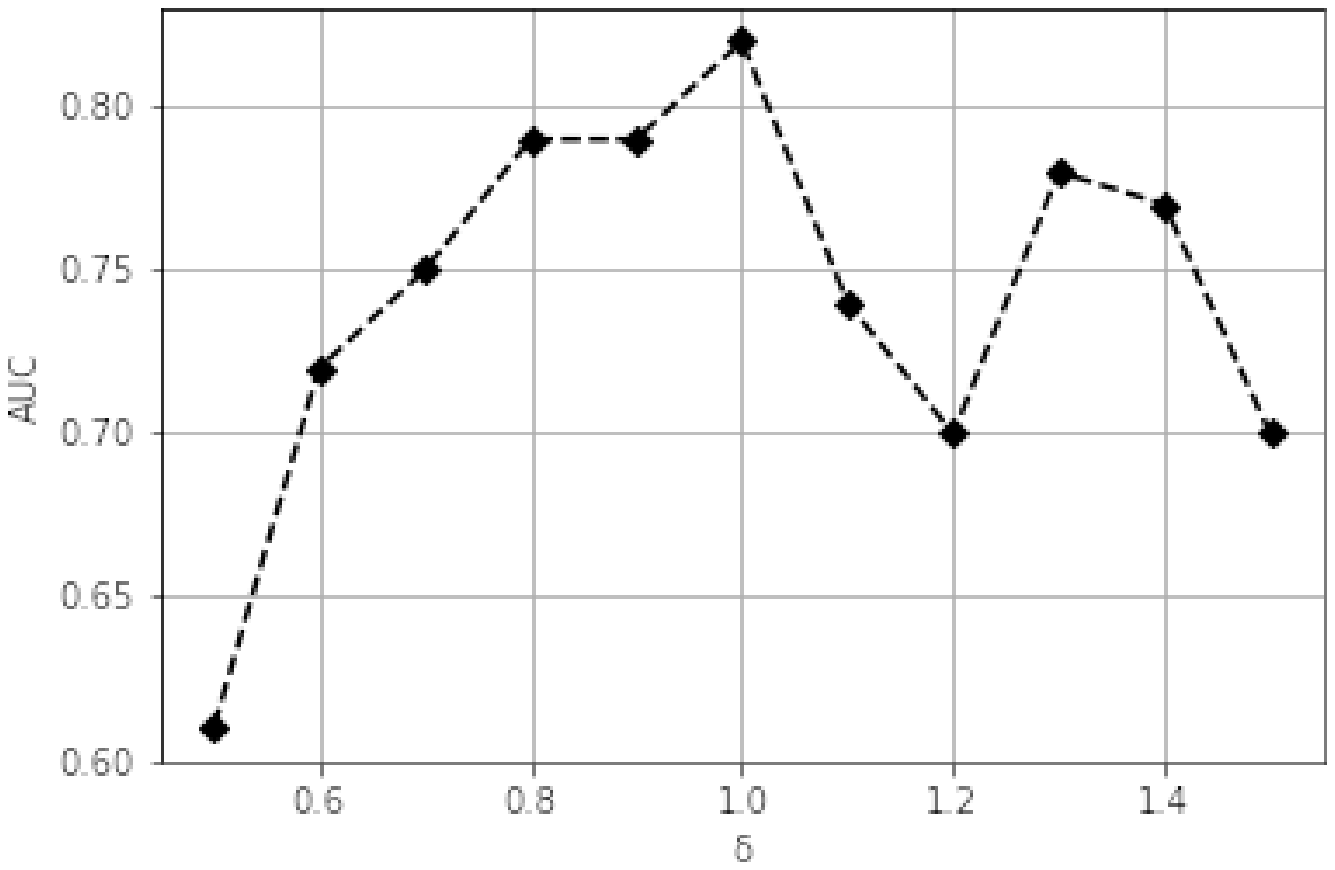} 
  \end{subfigure} 
  \caption{AUC metric for the GDNN when considered as a classifier for both the wave distorted case (a) and the upwelling distored case (b). On the x-axis the values of ZC parameters ($\delta$, $\delta_s$) and on the y-axis the AUC score.}
   \label{fig:GDN_as_clf} 
\end{figure}

As we can see in Fig.~ \ref{fig:GDN_as_clf}, the GDNN model appears to be less accurate than the CNN model. 
The reference case data show a lower AUC (compare to Fig. 3a) and  we can observe a general reduction of 0.1 
AUC respect to the results obtained with the CNN model. When feeding the GDNN model with ZC data with a 
different tuning of parameters $\delta$, we can observe that GDNN tends to be more degraded at $\delta <1$
then the CNN model (compare to Fig. 8a);  in fact, the AUC can lose up to 0.21 with respect to reference case. 
Note, that the same tuning of parameter $\delta$ would reveal a plateau in the AUC score whose values is much 
closer to that one attained in the reference case. When considering the distortion of parameter $\delta_s$ we 
can still appreciate a degradation at values lower than 0.3. However, the decrease in the AUC scores appears 
milder ($\sim 0.1$) respect to that shown for the CNN model. On the contrary, as $\delta_s > 0.3$ there is 
a significant reduction in the AUC scores; with  respect to the reference case the AUC scores can now be 
reduced up to 0.2.  

\section{Summary and Discussion}

This work was strongly motivated to understand  the high skill in ENSO prediction obtained with the CNN approach in 
\cite{Ham2019} in particular at long lead times. Although heat maps were presented in \cite{Ham2019}, their 
analysis does not connect immediately to the detailed processes of ENSO dynamics, which is also difficult 
because of the wide range of data they used.  In this paper, we introduced distorted physics simulations with 
the well-known Zebiak-Cane (ZC) model \cite[]{Zebiak1987} to determine how a CNN can perform on real 
data when trained  on data from `wrong' model simulations. 

The behavior of the ZC model can be elegantly described by a delay-differential equation 
\cite[]{Suarez1988, Jin1997a} 
\begin{eqnarray}
\frac{dT(t)}{dt} = a T(t) - b T(t-d) - c T^{3}(t)
\label{e:DA} 
\end{eqnarray}
for the eastern Pacific temperature $T$ as a function of time $t$. Here the constant $a$ 
indicates the strength of the positive feedbacks, $b$ that of the  delayed negative feedback 
(with a delay $d$ due to equatorial wave dynamics) and $c$ measures the strength of the 
nonlinear equilibration. 

By distorting the $\delta$ parameter in the ZC model, we modify the delay $d$ in (\ref{e:DA}) and hence 
mostly the adjustment processes in the equatorial Pacific. When the equatorial wave speeds
are distorted,  there is an asymmetry in the skill of the CNN. For faster waves $\delta < 1$, the 
performance remains good whereas for $\delta > 1$ (slower waves) it deteriorates. For example,
in case  $\delta= 1.2$ the  El Ni\~no event appears to be mainly constituted 
by slower oscillations, even though the behaviour of the  large-scale thermocline depth and sea 
surface temperature  is similar to the reference case. However, the loss of details on shorter time 
scales  leads the model  to still reasonably solve the classification task. 

By distorting the parameter $\delta_s$, we basically modify the feedback parameter $a$ in 
 (\ref{e:DA}) and hence the amplitude of the El Ni\~no events. However, also the stability 
 properties of the background climate state are changed as seen through the shift in the 
 Hopf bifurcation with $\delta_s$  \cite[]{vanderVaart2000TheModel}.  For increasing 
 $\delta_s$ and constant $\mu$ (as is done here), the background destabilizes as can 
 also be seen in Fig.~\ref{fig:ONI_distorted_delta_s}. The case $\delta_s= 0.1$ (reference 
 case $\delta_s = 0.3$)  offers a clear example about how the 
manipulation in the upwelling feedback   can degrade the AUC, i.e. the distortion of the  patterns in 
the data leads to a misplacement and  misalignment and reduce the capability of the  network in 
capturing the right patterns at the right (temporal)  location.  For other cases (e.g.  $\delta_s= 0.25$, 
$\delta_s= 0.45$ and $\delta_s= 0.6$)  the skill  of the CNN predictions is reduced less, because  the 
right combination of peaks and valleys in the  time series are present. Indeed, the absence of oscillating 
terms located at frequency band  4-8 months does not allow the CNN to capture all the relevant 
patterns  but only a part of them. 

The results indicate  that the accuracy of the   classification of the El Ni\~no and La Ni\~no  events for lead 
times of 9 months using a CNN approach is strongly related  to the capability of the CNN to capture the wave 
adjustment and feedback processes.   The exact combination of specific patterns like peaks 
and valleys occurring at specific regions of the time domain of all features  is  essential to generate
skill in the CNN  predictions.  The distorted physics approach can be very useful to look at how a 
CCN  based prediction scheme  can represent additional processes then considered here. The 
work of \cite{Ham2019}  has already  indicated that  connections between  the Indian-Pacific
\cite[]{Izumo2010}  and Atlantic-Pacific  \cite[]{Ham2013}   and extratropical-tropical  connections   
\cite[]{Zhao2020} are worth investigating. The latter 
interactions  have been described as ocean-atmosphere  meridional modes and can  influence 
ENSO and tropical variability on decadal time scales from both  hemispheres  independently 
\cite[]{Amaya2019}. However, one cannot do this with the Zebiak-Cane  model and needs 
to do such distorted physics simulations with a more sophisticated global climate model.

\section*{Conflict of Interest}
The authors declare to have no conflicts to disclose

\section*{Competing interests}
The authors declare to have no competing interest

\section*{Author Contributions}
All authors contributed to the design of this study. Results were mainly obtained by GL and IG. The paper was jointly written with contributions from all authors. 

\section*{Data Availability Statement}
The data that support the findings of
this study are openly available in github at
\url{https://github.com/glancia93/Physics-captured-by-data-based-methods-in-El-Nino-prediction\_PyCODE}

\section*{acknowledgments}
The work by HD was sponsored by the Netherlands Science Foundation (NWO) through the  project OCENW.M20.277. 

\bibliographystyle{apalike}  
\bibliography{cnn.bib, non_zotero_modified.bib, bib_plus.bib}

\end{document}